\documentclass[a4paper,fleqn,usenatbib]{mnras}
\usepackage[T1]{fontenc}
\usepackage{ae,aecompl}
\usepackage{graphicx}	
\usepackage{amsmath}	
\usepackage{amssymb}	
\usepackage{booktabs}	
\usepackage[normalem]{ulem}
\usepackage{overpic} 
\usepackage{url}

\usepackage{newtxtext,newtxmath}

\usepackage{tablefootnote,footnotehyper} 
\usepackage{array}

\newcolumntype{H}{>{\setbox0=\hbox\bgroup}c<{\egroup}@{}} 

\newcommand{\chron}{\mathcal{K}}

\title[Grid modelling of $\delta$\,Sct stars]{A grid of 200\,000 models of young $\delta$~Scuti stars using MESA and GYRE}

\author[Simon J. Murphy et al.]{
Simon J. Murphy,$^{1}$\thanks{E-mail: simon.murphy@usq.edu.au (SJM)} Timothy R. Bedding,$^{2}$ Anuj Gautam$^{1}$ and Meridith Joyce$^{3,4}$
\\
$^{1}$ Centre for Astrophysics, University of Southern Queensland, Toowoomba, QLD 4350, Australia\\
$^{2}$ Sydney Institute for Astronomy, School of Physics, University of Sydney NSW 2006, Australia\\
$^{3}$ Konkoly Observatory, Research Centre for Astronomy and Earth Sciences, H-1121 Budapest Konkoly Th. M. \'ut 15-17., Hungary\\
$^{4}$ CSFK, MTA Centre of Excellence, Budapest, Konkoly Thege Mikl\'os \'ut 15-17., H-1121, Hungary\\
}

\date{Accepted XXX. Received YYY; in original form ZZZ}
\pubyear{2021}
\begin{document}
\label{firstpage}
\pagerange{\pageref{firstpage}--\pageref{lastpage}}
\maketitle

\begin{abstract}
The rapidly increasing number of delta Scuti stars with regular patterns among their pulsation frequencies necessitates modelling tools to better understand the observations. Further, with a dozen identified modes per star, there is potential to make meaningful inferences on stellar structure using these young $\delta$\,Sct stars.
We compute and describe a grid of $>$200,000 stellar models from the early pre-main-sequence to roughly one third of the main-sequence lifetime, and calculate their pulsation frequencies.
From these, we also calculate asteroseismic parameters and explore how those parameters change with mass, age, and metal mass fraction. We show that the large frequency separation, $\Delta\nu$, is insensitive to mass at the zero-age main sequence. In the frequency regime observed, the $\Delta\nu$ we measure (from modes with $n\sim5$--9) differs from the solar scaling relation by $\sim$13\%. We find that the lowest radial order is often poorly modelled, perhaps indicating that the lower-order pressure modes contain further untapped potential for revealing the physics of the stellar interior.
We also show that different nuclear reaction networks available in {\sc mesa} can affect the pulsation frequencies of young $\delta$\,Sct stars by as much as 5\%. 
 We apply the grid to five newly modelled stars, including two pre-main-sequence stars each with 15+ modes identified, and we make the grid available as a community resource.
\end{abstract}

\begin{keywords}
asteroseismology -- stars: evolution -- stars: fundamental parameters -- stars: pre-main-sequence -- stars: variables: $\delta$ Scuti
\end{keywords}



\section{Introduction}
\label{sec:intro}

The discovery of regularly spaced pulsation frequencies amongst the pressure modes of young intermediate-mass (1.3--2.2\,M$_{\odot}$) stars \citep{beddingetal2020} has ushered in a new era of asteroseismic investigation. The once-thorny mode identification problem \citep{guziketal2021a,kurtz2022} is steadily gaining traction for some young delta Scuti stars. Asteroseismic ages from several $\delta$\,Sct stars in young clusters and/or stellar associations have now been determined \citep{murphyetal2021a,steindletal2022,kerretal2022a,kerretal2022b,murphyetal2022a,currieetal2023a,scuttetal2023}. The recent discovery of many new $\delta$\,Sct stars in TESS light curves of the Pleiades \citep{beddingetal2023} suggests that asteroseismic ages may come to sit beside isochrones \citep[e.g.][]{gagneetal2023}, kinematics \citep[e.g.][]{squicciarinietal2021,miret-roigetal2022,zerjaletal2023} and lithium depletion \citep[e.g.][]{galindo-guiletal2022,woodetal2023} as key methods for dating young clusters.

For solar-like oscillators, that regular frequency spacing is known as the large-frequency separation, $\Delta\nu$, and it scales with the square-root of the mean stellar density, $\Delta\nu \propto \sqrt{\rho}$ \citep{ulrich1986,kjeldsen&bedding1995}. This has been exploited extensively in the \textit{Kepler} era to characterise red giants \citep[see reviews by][]{hekker2020,basu&hekker2020,jackiewicz2021} and perform Galactic archaeology \citep[see review by][]{serenellietal2021}.
Models have predicted that a similar scaling relation would apply to $\delta$\,Sct stars \citep{reeseetal2008,suarezetal2014}. Using eclipsing binaries, \citet{garciahernandezetal2015,garciahernandez2017} verified that the $\Delta\nu$ determined at low radial orders correlated with stellar density, and that this scaling relation also applies to rotating stars. Inclusion of rotation in the models reduces the scatter in the derived scaling relation \citep{rodriguez-martinetal2020}. Even rapid rotators show regular frequency patterns in pulsation models \citep{mirouhetal2019}, although those patterns do become less discernible \citep{mirouh2022}. This is borne out in K2 observations of $\delta$\,Sct stars in the Pleiades, where even stars with $v\sin i \sim 200$km\,s$^{-1}$ have a measurable $\Delta\nu$ despite some modes being missing \citep{murphyetal2022a}.

Regular spacings appear to be limited to young $\delta$\,Sct stars for a few reasons: in all but the fastest rotators, the pulsation frequencies of pressure modes (p\:modes) and gravity modes (g\:modes) are well separated for young stars, but this separation decreases with age, causing the p and g\:modes to interact via avoided crossings, spoiling the patterns \citep{christensen-dalsgaard2000,lignieres&georgeot2009,aertsetal2010}. In addition, in stars nearer the terminal-age main sequence, nuclear burning will have produced sharp molecular weight boundaries at the edge of the convective core, hence sharp sound-speed gradients, which also spoil the regular patterns \citep{reeseetal2017,dornan&lovekin2022,wintheretal2023}. Young stars, on the other hand, have recently been fully convective and have not undergone much nuclear burning. Space photometry has verified that it tends to be the young $\delta$\,Sct stars, such as those in young associations, that have regular spacings (e.g. \citealt{beddingetal2020,kerretal2022a}), as opposed to the somewhat older field stars observed by \textit{Kepler}, hence we focus on young stars in this paper.

A common approach to asteroseismic modelling is to construct an $n+1$ dimensional grid of $n$ stellar parameters that describe evolutionary tracks, whose properties are evaluated along the additional time dimension \citep[e.g.][]{sanchezariasetal2017}. Such properties include `classical' observables such as temperature and luminosity, and asteroseismic properties, namely the stellar oscillation frequencies. In this paper we present such a grid, calculated for young $\delta$\,Sct stars, that has already been used to model various targets \citep{kerretal2022a,kerretal2022b,murphyetal2022a,currieetal2023a,scuttetal2023}. We also make this grid available as a community resource.
We describe the physics and computation of the grid, as well as calculation of the pulsation frequencies, in Sec.\,\ref{sec:models}. We explore various asteroseismic parameters across the grid in Sec.\,\ref{sec:analysis}, including scaling relations in Sec.\,\ref{ssec:scaling}. In Sec.\,\ref{sec:real} we apply this grid to real stars, including two new pre-main-sequence (`pre-MS') stars that have never been modelled asteroseismically.


\section{Modelling parameters}
\label{sec:models}

\subsection{MESA evolutionary models}

\begin{figure}
\begin{center}
\includegraphics[width=0.48\textwidth]{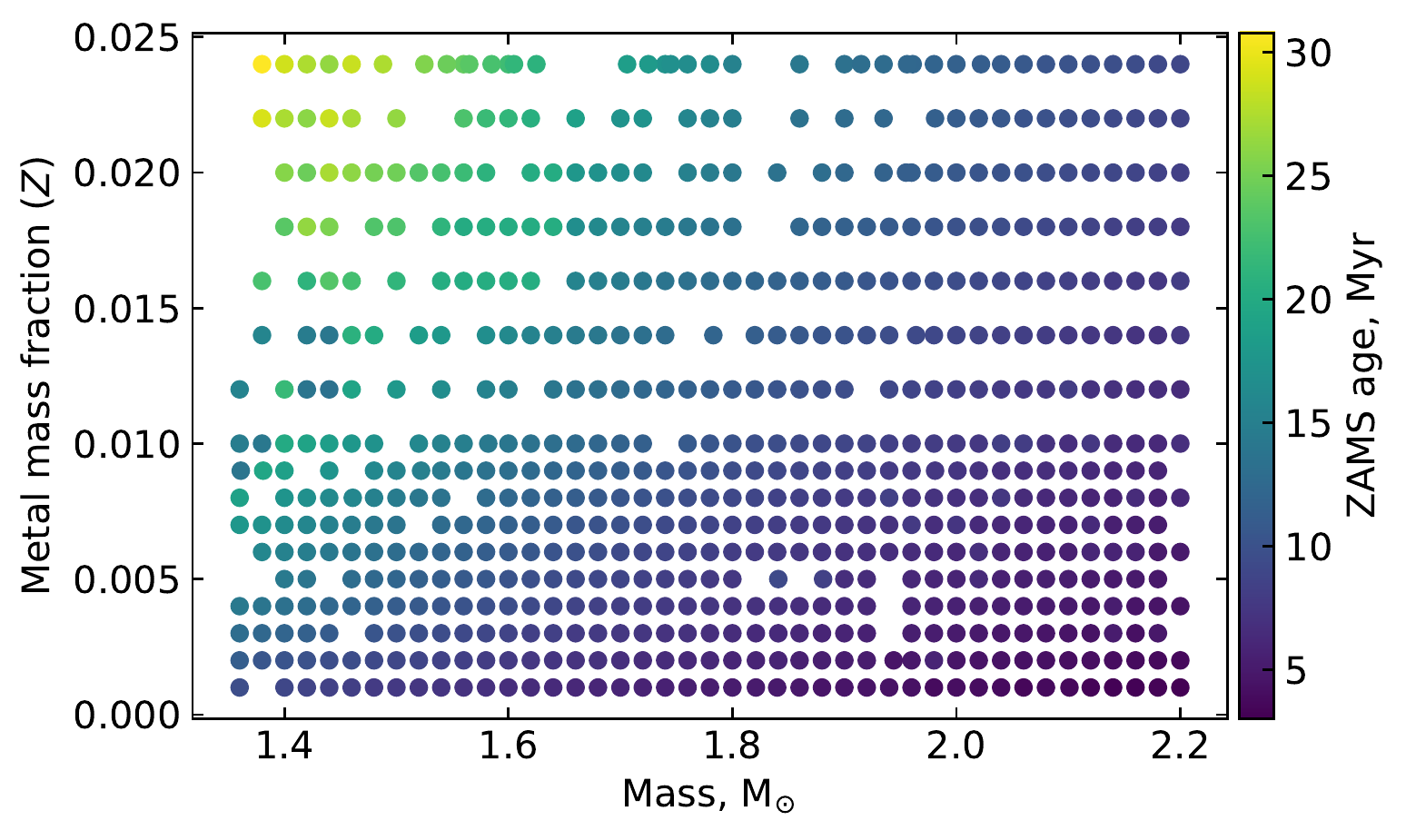}
\caption{Distribution of tracks across a mass--metallicity grid. Shown in colour is the stellar age at the ZAMS, as defined in Sec.\,\ref{ssec:Dnu}.}
\label{fig:mz_grid}
\end{center}
\end{figure}

Stellar evolutionary models were calculated with {\sc mesa} r15140 \citep{paxtonetal2011,paxtonetal2013,paxtonetal2015,paxtonetal2018,paxtonetal2019}. We calculated pre-MS models using an initial core temperature of $9\times10^{5}$\,K, which determines age=0 for our models, and evolved them to approximately one third of their MS lifetimes (for reasons outlined in Sec.\,\ref{sec:intro}). The independent variables of the models were mass and metallicity; other `variables' were either dependent or fixed. Neither rotation nor accretion was included in these models. A representative {\sc mesa} inlist is provided.

Models were calculated in a grid that is approximately uniformly spaced in mass, from 1.36 to 2.20\,M$_{\odot}$ at 0.02-M$_{\odot}$ spacing. Spacing in metallicity is exact, with initial metal mass fractions, $Z$, between 0.001 and 0.010 at a spacing of 0.001, then spaced by 0.002 from 0.012 to 0.024. However, not all {\sc mesa} tracks ($M$--$Z$ pairs) successfully converged, largely due to idiosyncrasies of the {\sc mesa} {\tt hydro solver} that have been fixed in later versions. For tracks that did not converge, the mass was increased by 0.001\,M$_{\odot}$ and the track was retried for up to five iterations before the track was abandoned. Hence, there is some heterogeneity in mass, as shown in Fig.\,\ref{fig:mz_grid}. The grid comprises 664 tracks.

\subsubsection{Composition and nuclear reactions}
\label{sssec:composition}

The initial mass fractions of hydrogen, $X$, and helium, $Y$, were calculated based on the independent variable $Z$. While \citet{steindletal2022} showed that helium abundance contributes to the uncertainty in modelling of young $\delta$\,Sct stars, \citet{murphyetal2022a} found that the effect on stellar densities is 1--2 orders of magnitude less than that of rotation. Given that our models do not include rotation, we chose not to set the helium abundance as an independent variable in our non-rotating models. Instead, we calculated helium abundances as a function of the stellar metallicity, as follows. For a given track, we first calculated the difference between the stellar metallicity and the solar metallicity, $dZ = Z - Z_{\odot}$, where for $Z_{\odot}$ we used the bulk solar metal mass fraction of 0.0142 from \citet{asplundetal2009}. The helium abundances were then calculated assuming a helium enrichment rate $dY/dZ = 1.4$, which is a relatively poorly constrained quantity in the literature but for which a value of 1.4 appears reasonable (\citealt{brogaardetal2012,tlietal2018,vermaetal2019,lyttleetal2021} and references therein). We adopted a helium mass fraction of $Y_0=0.28$, and used our adopted helium enrichment ratio to calculate the stellar helium mass fraction:
\begin{eqnarray}
Y = Y_0 + dZ \frac{dY}{dZ}. \label{eq:dydz}
\end{eqnarray}
The hydrogen mass fraction constitutes the remainder of the initial composition, with
\begin{eqnarray}
X + Y + Z = 1.
\end{eqnarray}

We included initial quantities of $^2$H (deuterium) equal to $2\times10^{-5}$ of $^1$H \citep{stahleretal1980,linsky1998}, and of $^3$He equal to $1.66\times10^{-4}$ of $^4$He. Stellar abundances otherwise followed \citet{asplundetal2009}, with the corresponding {\tt A09} opacity settings (readers are encouraged to inspect the {\tt inlist} supplied). We used the {\tt jina reaclib} reaction rates \citep{cyburtetal2010} and the {\tt pp\_and\_cno\_extras} nuclear reaction network. This is the most complete nuclear reaction network relevant to the stars in our grid, but it comes at a greater computational cost. We compare the effects of different nuclear reaction networks on the stellar evolution (hence on mode frequencies) against their computation times in Appendix~\ref{app:nuclear}. We note the main conclusion here: that {\tt basic.net} is inadequate for modelling $\delta$\,Sct stars of all ages, producing frequency errors up to 5\%, and that both {\tt hot\_cno} and {\tt pp\_extras} are required. Combining these two networks with {\tt basic.net} will provide all the performance of {\tt pp\_and\_cno\_extras.net} but reduce computation time by 10\% compared to using {\tt pp\_and\_cno\_extras.net} directly.

\subsubsection{Mixing, convection, and atmospheres}
\label{sssec:mixing}

Stars are fully convective for part of the pre-MS stage, hence have a uniform composition to which different mixing parameters make little difference. Perhaps for this reason, \citet{murphyetal2021a} found that the chosen value of $\alpha_{\rm MLT}$ was unimportant to their asteroseismic fitting (see also \citealt{joyce&tayar2023}). Since we are interested in young stars, we followed that result, fixing $\alpha_{\rm MLT}$ to 1.9 for all tracks. We adopted the \citet{henyeyetal1965} formalism for the mixing length, and we did not include element diffusion or thermohaline mixing.

Convective overshoot becomes increasingly important for A-type stars as they approach the terminal-age main-sequence (TAMS), but the amount of convective core overshooting required is not well established and may depend on many stellar parameters \citep{lovekin&guzik2017,claret&torres2018,johnston2021,dornan&lovekin2022}. For the models in this work, the important convection zones are a thin one at the surface, and the convective core, each of which should have its own overshooting parameters.
\citet{pedersenetal2018} discussed appropriate values for the terms $f$ and $f_0$ for each zone and we adopted the values from \citet{pedersenetal2021} in this work. To be specific, we included exponential overshooting at the top of the hydrogen-burning core, with $f=0.022$ and $f_0=0.002$, and we included exponential overshooting in any non-burning shell (i.e. the stellar surface) with $f=0.006$ and $f_0=0.001$. Further description of these terms can be found in the {\sc mesa} documentation.\footnote{\url{https://docs.mesastar.org/en/r15140/index.html}} We note that the form of overshooting can be just as important as the magnitude \citep{anders&pedersen2023}, but we leave experiments of the effects of this to future work, especially since \citet{sanchezariasetal2017} found that different overshooting prescriptions matter little to the p-mode frequencies of $\delta$\,Sct stars. That result, and the fact that we terminated the evolution one third of the way through the main sequence, suggested that convection parameters are somewhat less important than in many other places on the HR diagram, hence we decided not to vary them. A systematic study of the effect of different convection treatments on the mode frequencies in $\delta$\,Sct models will be the subject of future work.

There is some debate on the importance of different stellar atmosphere treatments. \citet{murphyetal2021a} tried four and found that none affected their asteroseismic age for the pre-MS star HD\,139614, at even the $1\sigma$ level. Conversely, \citet{steindletal2021b} found Eddington--Gray atmospheres were preferred. Atmosphere prescriptions in {\sc mesa} have changed since both of those papers. In this work with r15140, we used `fixed' Eddington T--$\tau$ atmospheres.

\subsubsection{Sampling the evolutionary tracks}
\label{sssec:sampling}

The evolution was broken into different {\sc mesa} inlists so that sampling parameters could be changed as needed. It is important to keep each time interval in the calculation small so as to minimise computation errors. Until an age of 1\,Myr, the evolution was calculated at intervals of 13\,kyr, with 15 intervals to each saved sample for a spacing of 0.2\,Myr between samples. In order to sample the rapid changes in evolution (and hence, pulsation frequencies) on the pre-MS, intervals between saved samples were then reduced to every 0.05\,Myr until an age of 10.5\,Myr. The subsequent evolution is slower and the sampling rate was decreased to every 3\,Myr until 40\,Myr. Thereafter, it was limited by changes in position on the HR diagram ($\Delta\log T_{\rm eff}=0.0006$ and $\Delta\log L=0.002$), with an upper limit of 100\,Myr between samples. The sampling was therefore mass and metallicity dependent. 

Spatially, zoning within the star was determined automatically and adaptively with {\sc mesa}. We used \mbox{{\tt mesh\_delta\_coeff} = 1.25} (larger meshes result in fewer zones and worse spatial resolution), which is larger than the default value of 1.0. The frequency error arising from this larger mesh is $<0.2$\%, but the computations are 20\% faster. A detailed characterisation of this is being prepared as part of our investigations into systematic modelling uncertainties (Gautam et al., in prep.).

Having created these models, the next step was to calculate pulsation frequencies.

\subsection{GYRE pulsation calculations}
\label{ssec:gyre}

\begin{figure}
\begin{center}
\includegraphics[width=0.48\textwidth]{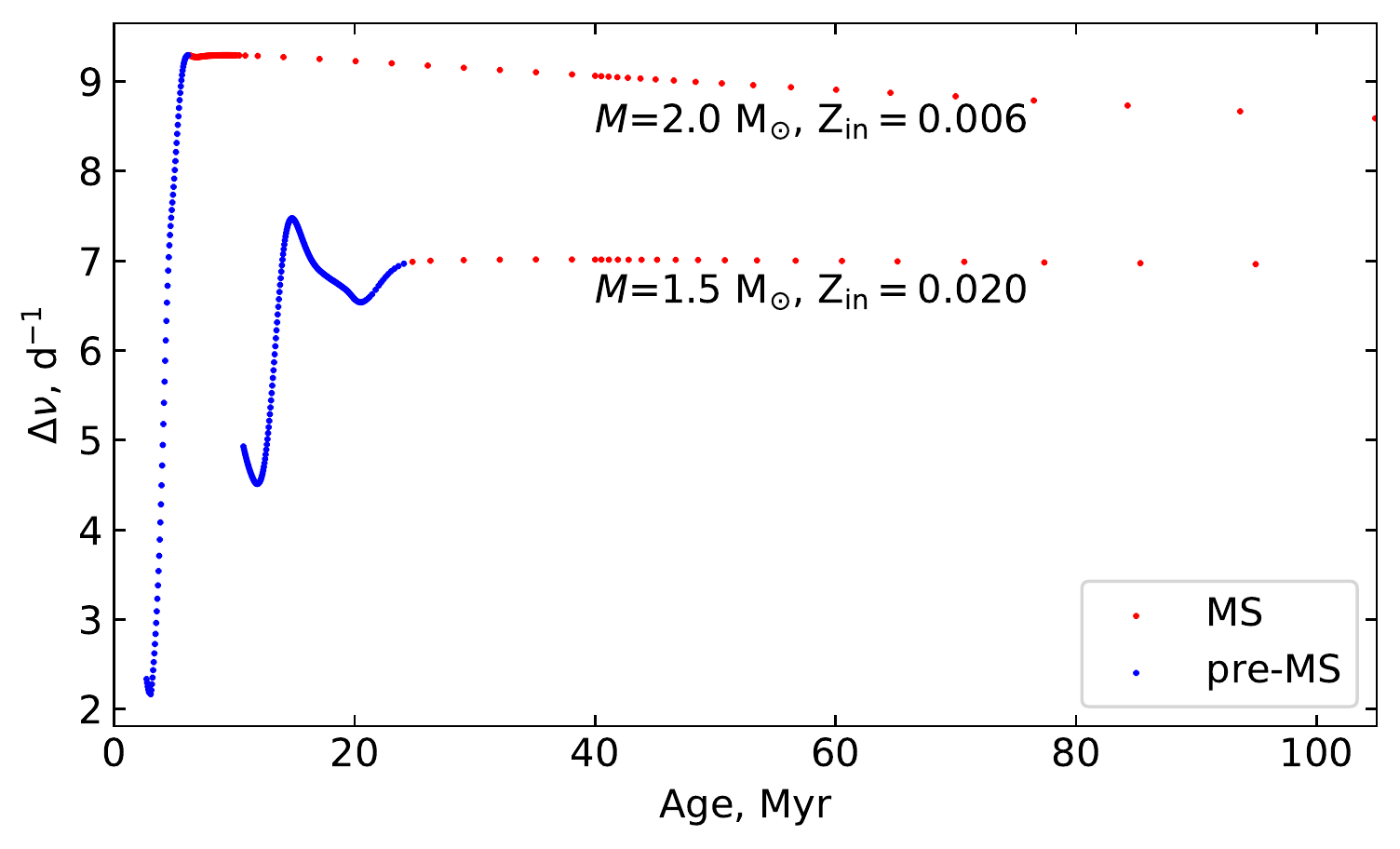}
\caption{Evolution of $\Delta\nu$ on the pre-MS and early MS, showing the dip that precedes the maximum $\Delta\nu$ value, and the empirical determination of the ZAMS defined in Sec.\,\ref{ssec:Dnu}. At 40\,Myr in both tracks, the switch between age-based sampling and HR-diagram positional sampling can also be seen (Sec.\,\ref{sssec:sampling}).}
\label{fig:ZAMS}
\end{center}
\end{figure}

Stellar pulsation frequencies were calculated with {\sc gyre} v6.0.1 \citep{townsend&teitler2013,GYREdoc}. We initially calculated these for a representative set of models across the ($M$, $Z$) parameter space at all ages. This allowed us to reduce the computational cost of the full grid by ignoring models that did not have $\Delta\nu$ in the region of interest ($\gtrsim5$\,d$^{-1}$). Specifically, we found that pre-MS stars have a dip in $\Delta\nu$ before reaching their maximum values (Fig.\,\ref{fig:ZAMS}), and we calculated pulsation frequencies from ages slightly before this ($M$-, $Z$-dependent) age.
We used dynamical limits for the frequency range over which modes were calculated, from 1.5$\times$ to 12$\times$ $\Delta\nu$. Since oscillation frequencies were not available prior to this calculation, $\Delta\nu$ was calculated using the {\sc mesa} value, which differs a little \citep{murphyetal2021a} from the $\Delta\nu$ we ultimately calculate with {\sc gyre} (for reasons stated in Sec.\,\ref{ssec:Dnu}). 

The {\sc gyre} calculations used the second-order Gauss-Legendre Magnus `{\tt MAGNUS\_GL2}' difference equation scheme \citep{kiehl1994,gander&vandewalle2007,GYREdoc}. Our tests of difference equation schemes revealed only small differences in the accuracy of these second-order calculations versus higher orders ($\delta f/f<2.5\times10^{-5}$, corresponding to <0.002\,d$^{-1}$ at 75\,d$^{-1}$), while fourth and sixth-order calculations take 30 and 60\% longer to compute, respectively. In fact, unlike in models of red-giant stars \citep{ylietal2023}, none of the difference equation schemes available in {\sc gyre} v6.0.1 resulted in appreciable differences in mode frequencies for these young $\delta$\,Sct stars, except for {\tt colloc2}, which has $\delta f/f\sim2.5\times10^{-4}$ for pre-MS stars. On the MS, {\tt colloc2} performs as well as {\tt MAGNUS\_GL2} and halves the computation time. We will further investigate the impact of {\sc gyre} computation parameters in future work.

We performed a linear scan over the described frequency range for up to 100 frequencies -- more than sufficient to capture all expected p\:modes, even if there are many g\:modes in the same range. After some experimentation to determine optimum scan parameters to preserve accuracy whilst saving computation time, the following scan parameters were used: {\tt x\_i} = 0.00001, {\tt w\_osc} = 10, {\tt w\_exp} = 2, and {\tt w\_ctr} = 10. These values are also the values recommended in the {\sc gyre} documentation. Other parameters were kept as their defaults, as shown in the {\tt gyre\_template} file provided.

For each model we calculated adiabatic p-mode frequencies for radial and dipole modes over radial orders $n\sim1$--11, and the occasional g-mode frequency that fell in our calculation range. Since the models are non-rotating, we took $m=0$ for all modes. We also ignored the g\:modes in the rest of this work.

\subsection{Asteroseismic parameters}
\label{ssec:Dnu}

Reliably inferring the asteroseismic large spacing, $\Delta\nu$, is extremely useful for two reasons: (i) Once $\Delta\nu$ is established, mode identification becomes much easier, because modes of a given degree align vertically in the \'echelle diagram \citep{beddingetal2020}; and (ii) the relation of $\Delta\nu$ to the square root of the mean stellar density tightly constrains age and metallicity.

The other asteroseismic parameter useful in mode identification and model characterisation is $\epsilon$, which parametrizes the positions of different ridges in the \'echelle diagram. In the asymptotic regime, the stellar oscillation frequencies can be expressed as
\begin{eqnarray}
\nu = \Delta\nu(n+\ell/2+\epsilon).\label{eq:deltanu}
\end{eqnarray}
While $\delta$\,Sct stars do not oscillate in the asymptotic regime, they do show equidistantly spaced frequencies (i.e. a large separation) at moderate values of $n\sim5$--9 \citep{beddingetal2020}, and with a $\Delta\nu$ slightly smaller than the truly asymptotic value (hence smaller than the {\sc MESA} value, \citealt{murphyetal2021a}). We used eq.\,\ref{eq:deltanu} to determine $\Delta\nu$ for our model frequencies. Specifically, we fitted a straight line to the radial mode frequencies from $n=5$ to 9 (inclusive) using linear regression. The resulting gradient is $\Delta\nu$ and the y-intercept is $\epsilon$ \citep{whiteetal2011c}. \citet{beddingetal2020} showed $\epsilon$ and $\Delta\nu$ to be informative for mass and age estimates on the MS, but pre-MS stars were not discussed. We analyse both evolutionary stages in Sec.\,\ref{sec:analysis}.

To evaluate trends in pulsation properties occurring at different evolutionary stages, we needed to distinguish pre-MS and MS models. While it is possible to define these stages in terms of core physics (e.g. the point after the CN equilibrium burning bump when nuclear burning accounts for $\geq1$\% of the total luminosity; \citealt{zwintzetal2014b}), we developed an empirical definition attuned to the pulsation properties by evaluating the evolutionary change in $\Delta\nu$, looking for it to flatten off to its MS value. We first determined the maximum value of $\Delta\nu$ on the MS by applying an age threshold of 30\,Myr, and we defined the ZAMS as the first time that the star reaches 95\% of this maximum (MS) $\Delta\nu$ and where $0<d\Delta\nu/dt<0.5$\,d$^{-1}$/Myr. For stars less massive than 1.65\,M$_{\odot}$ having $Z>Z _{\odot}$, we also imposed a 20\,Myr minimum on the ZAMS age to avoid confusion with pre-MS excursions in $\Delta\nu$, as shown in Fig.\,\ref{fig:ZAMS}. 

The faster rate at which $\Delta\nu$ changes during the pre-MS necessitated the finer sampling during the early evolution described in Sec.\,\ref{sssec:sampling}. The grid comprises 136\,586 pre-MS and 106\,665 MS models. The number of models available in each stage therefore does not correspond to the duration of those stages. Hence, for stellar parameter estimation via asteroseismology (Sec.\,\ref{sec:real}), we used the neural network described in \citet{scuttetal2023} that was trained on this grid. 

\begin{figure*}
\begin{center}
\includegraphics[width=0.98\textwidth]{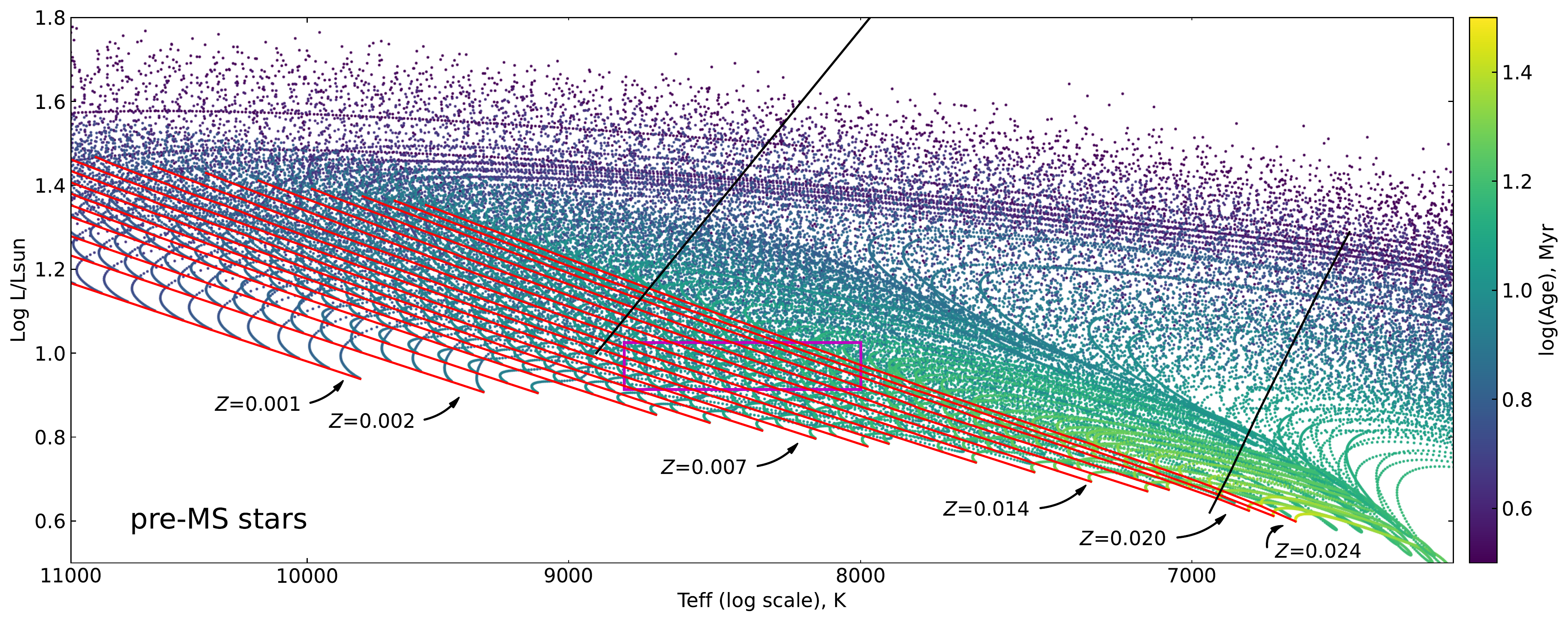}\\
\includegraphics[width=0.98\textwidth]{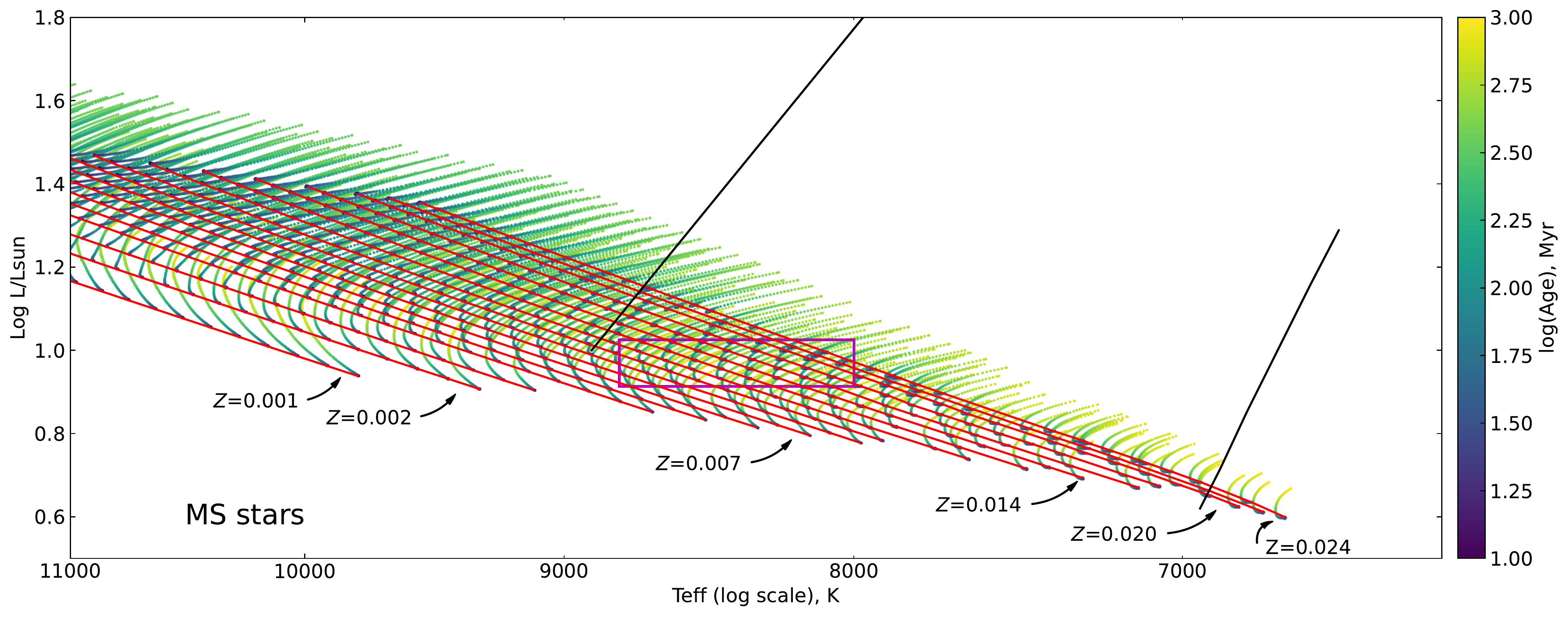}\\
\includegraphics[width=0.98\textwidth]{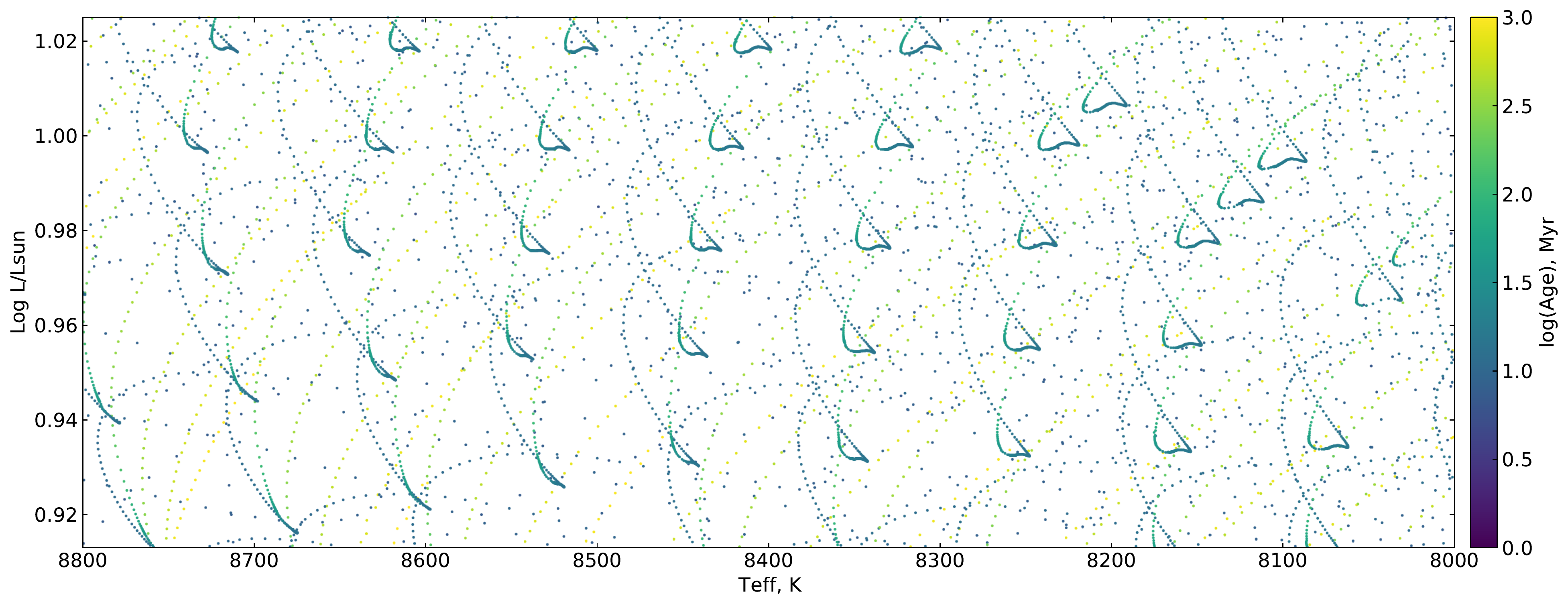}
\caption{Top: All pre-MS tracks with M>1.45\,M$_{\odot}$, colour-coded by age. The diagonal red lines show the ZAMS labelled by metal mass fraction. The solid black lines are the theoretical instability strip boundaries for solar metallicity from \citet{dupretetal2004}. The magenta box shows a typical 2$\sigma$ uncertainty for an observed $\delta$\,Sct star. Middle: as top, but for MS tracks. Bottom: Zoom at the centre of the instability strip, corresponding to the magenta box in other panels, showing the wide variety of tracks that can represent the typical observed $\delta$\,Sct star at the 2$\sigma$ level.}
\label{fig:HR_ZAMS}
\end{center}
\end{figure*}

We show the grid on an HR diagram in Fig.\,\ref{fig:HR_ZAMS}a. Even without rotation, somewhat different tracks can overlap in the HR diagram (Fig.\,\ref{fig:HR_ZAMS}b), explaining why some stars with similar atmospheric parameters can exhibit such different pulsation spectra \citep{balona2014}.

\subsection{Description of the grid data file}
\label{ssec:grid-export}

We make the grid available as a community resource as a csv file. The grid file contains 243\,252 rows, including the column-header row. Descriptions of the columns are provided in the accompanying readme file. All saved {\sc mesa} profiles are present in the csv file, sorted by evolutionary track -- the tracks themselves are sorted by mass, then metallicity. The asteroseismic parameters $\Delta\nu$ and $\epsilon$ are available for the 190\,331 models on which we ran {\sc gyre} (see Sec.\,\ref{ssec:gyre}).


\section{Analysis}
\label{sec:analysis}

\subsection{The behaviour of $\Delta\nu$ and $\epsilon$}
\label{ssec:behaviour}

\begin{figure}
\begin{center}
\includegraphics[width=0.48\textwidth]{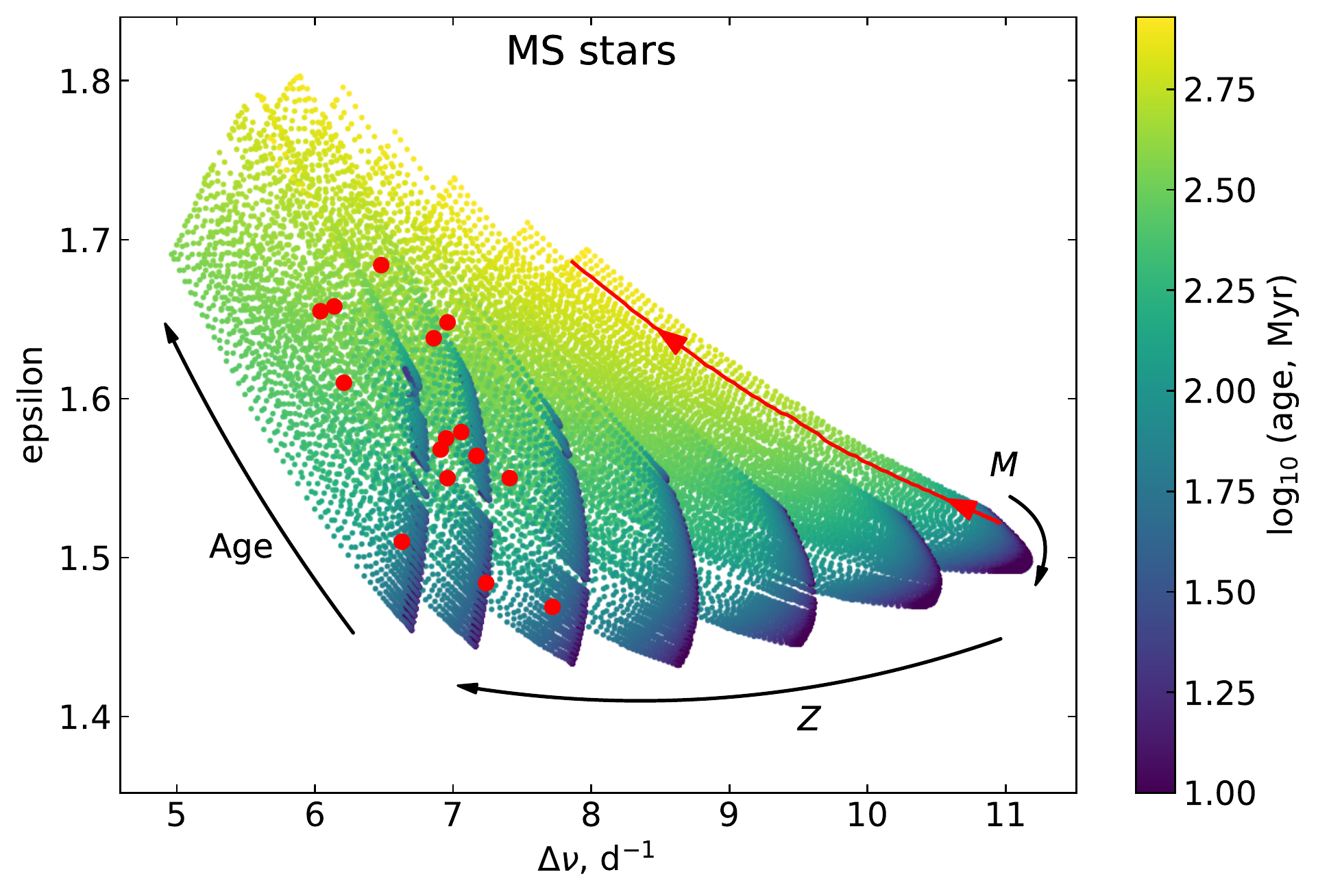}
\caption{Evolution of $\Delta\nu$ and $\epsilon$ on the main sequence, showing dependence on mass, metallicity, and age in a predictable, monotonic way. Models shown here have $M\geq1.53$\,M$_{\odot}$ and $Z\geq0.002$. Tracks have been thinned in metallicity for clarity. The evolutionary track at $M=1.60$\,M$_{\odot}$ and $Z=0.002$ is shown as a red line. The 15 stars from Table\:\ref{tab:nature} are shown as red circles.}
\label{fig:dnu-eps-ms}
\end{center}
\end{figure}

Even without individual frequency modelling, parametrization of ridges in the \'echelle diagram can yield useful information about the star. In Fig.\,\ref{fig:dnu-eps-ms}, we plot the evolution of $\Delta\nu$ and $\epsilon$ on the MS as a function of the input parameters, $M$ and $Z$. To this figure, we add the stars from \citet[][their extended data figure 2]{beddingetal2020}, after revisiting all of their \'echelles to re-determine $\Delta\nu$, $\epsilon$, and $f_1$ (we provide these in Table\,\ref{tab:nature}). We dropped the three stars whose \'echelles we were unable to model due to insufficiently clear ridges.

\begin{table}
\begin{center}
\caption{Asteroseismic quantities for 15 stars from extended data figure 2 of \citet{beddingetal2020}. We remeasured the asteroseismic large spacing $\Delta\nu$, the phase parameter $\epsilon$, and the fundamental radial mode frequency $f_1$ in this work.}
\begin{tabular}{r r c c c r@{ $\pm$ }l}
\toprule
	\multicolumn{1}{c}{TIC}	&	\multicolumn{1}{c}{HD}	&	$\Delta\nu$	 &	$\epsilon$	&	$f_1$	&	\multicolumn{2}{c}{$v \sin i$} \\
		&		&	d$^{-1}$ & & d$^{-1}$ &	\multicolumn{2}{c}{km\,s$^{-1}$} \\
\midrule
$	9147509	$&$	25369	$&$	6.14	$&$	1.658	$&$	18.96	$&	\multicolumn{2}{c}{}			\\
$	11361473	$&$	290799	$&$	7.72	$&$	1.469	$&$	22.53	$&\multicolumn{2}{c}{}		\\
$	34737955	$&$	44930	$&$	6.04	$&$	1.655	$&$	18.44	$&	\multicolumn{2}{c}{}			\\
$	43363194	$&$	3622	$&$	6.86	$&$	1.638	$&$	20.39	$&$	50	$&$	6	$\\
$	44645679	$&$	24975	$&$	6.21	$&$	1.610	$&$	18.60	$&$	88	$&$	4	$\\
$	172193026	$&$	46722	$&$	6.48	$&$	1.684	$&$	20.23	$&\multicolumn{2}{c}{}			\\
$	255548143	$&$	44958	$&$	6.96	$&$	1.550	$&$	20.57	$&$	114	$&$	11	$\\
$	259675399	$&$	31640	$&$	6.63	$&$	1.510	$&$	19.95	$&$	136	$&$	4	$\\
$	272951803	$&$	187547	$&$	6.96	$&$	1.648	$&$	21.71	$&$	10	$&$	2	$\\
$	274038922	$&$	20203	$&$	7.41	$&$	1.550	$&$	21.74	$&$	40	$&$	25	$\\
$	287347434	$&$	99506	$&$	7.06	$&$	1.579	$&$	21.19	$&$	26	$&$	2	$\\
$	294157254	$&$	55863	$&$	6.91	$&$	1.568	$&$	20.75	$&$	99	$&$	5	$\\
$	316920092	$&$	31901	$&$	6.95	$&$	1.575	$&$	21.07	$&$	33	$&$	4	$\\
$	388351327	$&$	70510	$&$	7.24	$&$	1.484	$&$	21.68	$&$	94	$&$	10	$\\
$	408906554	$&$	42005	$&$	7.17	$&$	1.564	$&$	21.55	$&$	130	$&$	30	$\\
\bottomrule
\end{tabular} 
\label{tab:nature}
\end{center}
\end{table}

We find that $Z$ affects $\Delta\nu$ strongly but has little effect on $\epsilon$: the arrow marking the $Z$ dependence in Fig.\,\ref{fig:dnu-eps-ms} lies almost parallel to the $\Delta\nu$ axis. Conversely, mass has little effect on $\Delta\nu$ but substantially more on $\epsilon$. We discuss this further in Sec.\,\ref{ssec:zams_dnu}. Thus, mass and metallicity are almost orthogonal in this plane, which is why $\Delta\nu$ and $\epsilon$ are useful asteroseismic parameters. There is also an age dependence whose vector lies at an angle to that of the other parameters. The simplicity of Fig.\,\ref{fig:dnu-eps-ms} suggests that machine learning based solely on these asteroseismic observables should perform quite well for MS stars (S.~Kumar~Panda, submitted). The inclusion of a temperature variable would also offer greater sensitivity to mass.

\begin{figure}
\begin{center}
\includegraphics[width=0.48\textwidth]{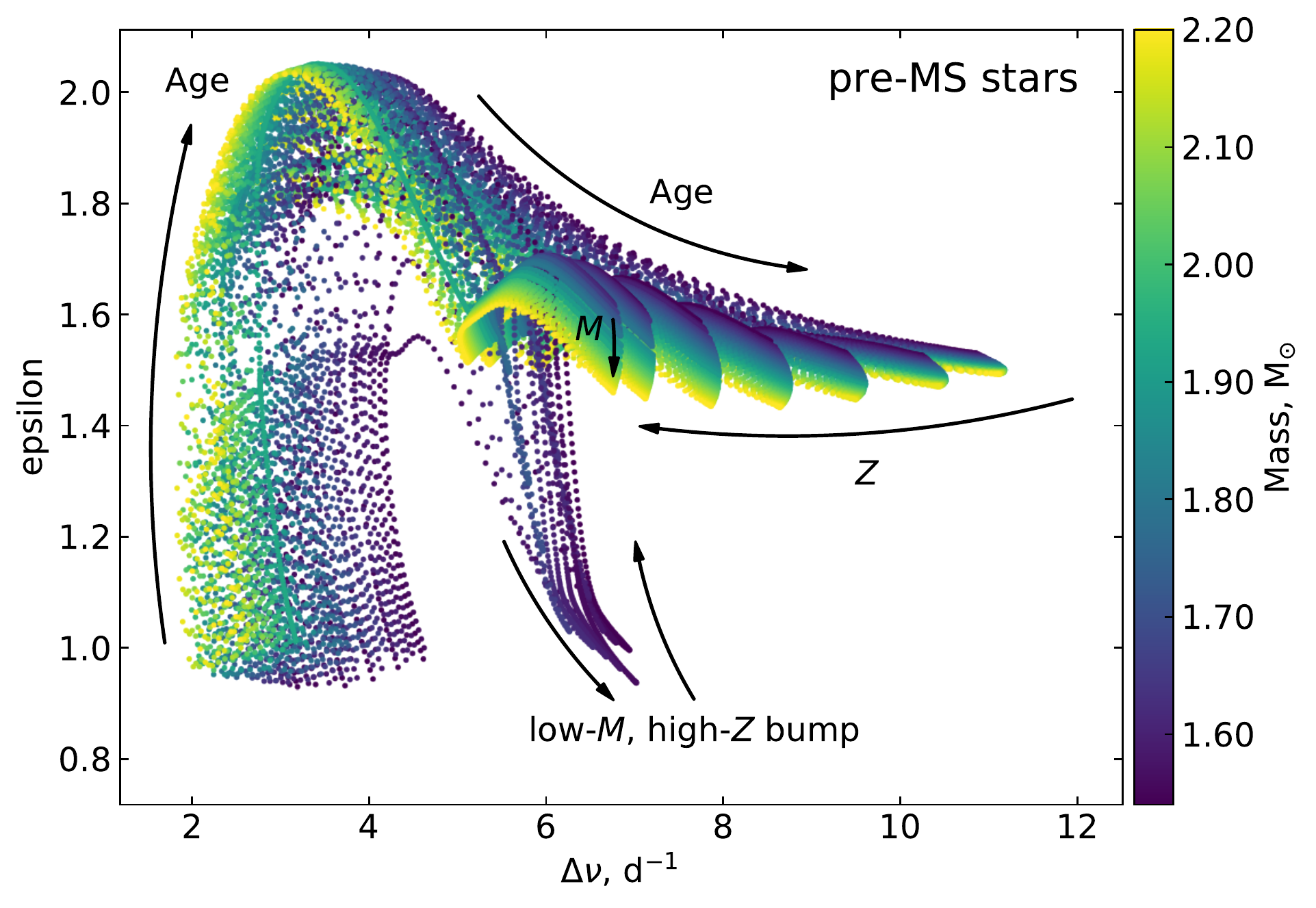}
\includegraphics[width=0.48\textwidth]{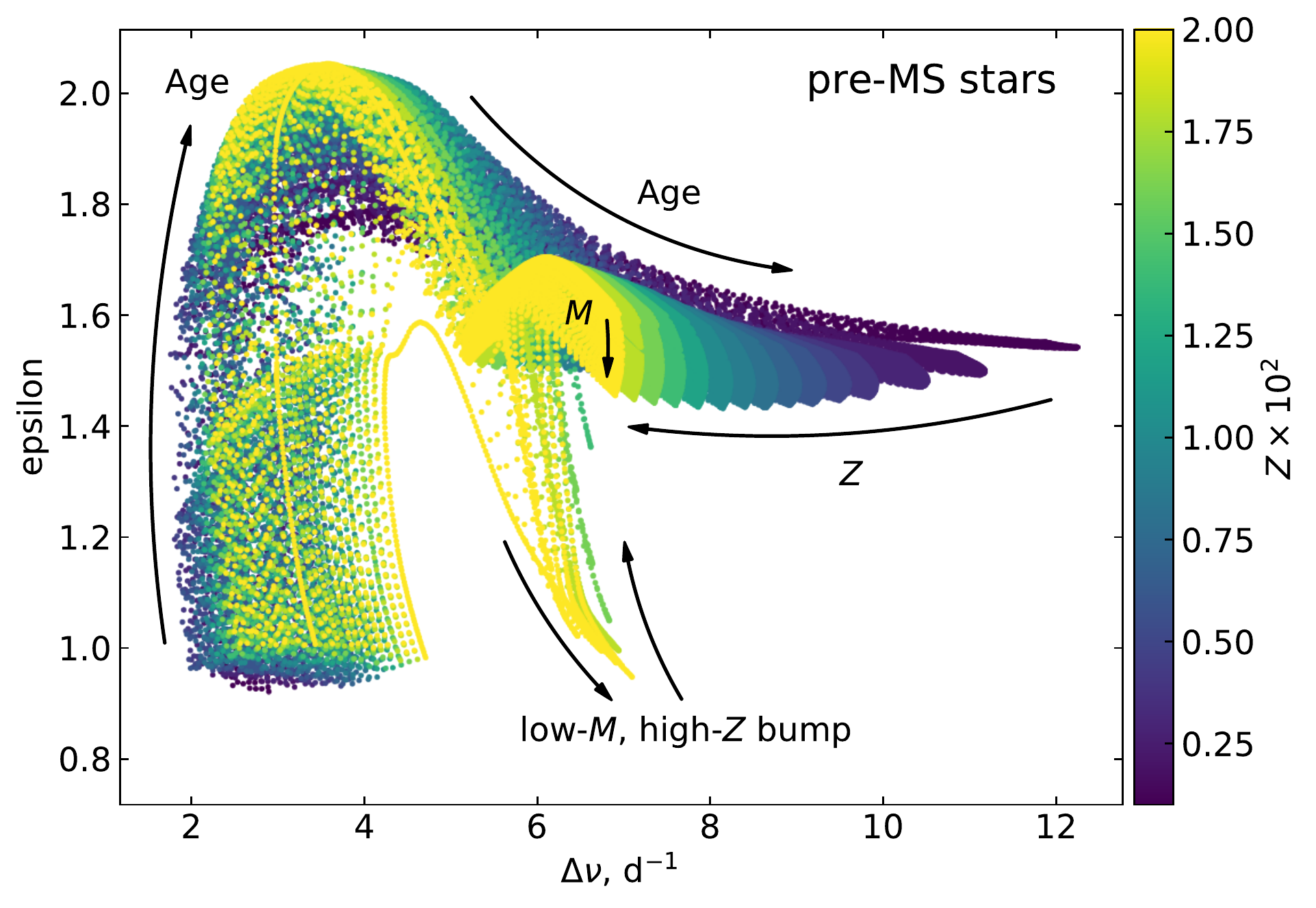}
\caption{The evolution of $\Delta\nu$ and $\epsilon$ on the pre-MS, showing dependence on mass, metallicity, and age. Models in the top panel are thinned like in Fig.\,\ref{fig:dnu-eps-ms} and colour-coded by mass, whereas in the bottom panel all values of $Z\leq0.02$ are shown and colour coding is by metallicity. This upper limit allows the low-$M$ high-$Z$ bump to be seen clearly (see Sec.\,\ref{ssec:behaviour}). Arrows on this bump show the direction of increasing age.}
\label{fig:dnu-eps-prems}
\end{center}
\end{figure}

The same diagram is somewhat more complicated for pre-MS stars, where the dependence on age is no longer monotonic for either parameter (Fig.\,\ref{fig:dnu-eps-prems}). At the youngest ages of a given evolutionary track ($M$--$Z$ pair), $\Delta\nu$ changes little with evolution while $\epsilon$ grows rapidly from the minimum to the maximum value of its observed range. A turning point is soon encountered, though, after which age and $Z$ share importance in determining $\Delta\nu$. To further complicate matters, low-mass and high-metallicity tracks behave differently from most others in the grid, undergoing an excursion to very low values of $\epsilon$. 

Although the range in $\epsilon$ is greater for the pre-MS models, this is really just for the initial contraction, where $\Delta\nu$ is small. The majority of the pre-MS stage is contained within a dense region of points having $\Delta\nu \sim 5$--10 and $\epsilon \sim 1.4$--1.7. A comparison of Figs\,\ref{fig:dnu-eps-ms} \& \ref{fig:dnu-eps-prems} shows that this is the same region occupied by the MS models. This is somewhat expected, given that the MS evolutionary tracks evolve back in the direction from whence they came on the pre-MS (Fig.\,\ref{fig:HR_ZAMS}), but it makes distinguishing MS and pre-MS stars difficult when only $\Delta\nu$ and $\epsilon$ are used. The full set of pulsation frequencies, however, is able to distinguish these two stages in the vast majority of cases, and a neural network is still able to learn to emulate the models either side of the ZAMS \citep{scuttetal2023}.

\subsection{Scaling Relations}
\label{ssec:scaling}

For solar-like oscillations, which are excited stochastically by convection, there are two widely-used scaling relations (see review by \citealt{hekker2020}). The first concerns $\nu_{\rm max}$, the frequency of maximum oscillation power, which is observed to scale as $g/\sqrt{T_{\rm eff}}$. 

The modes of $\delta$\,Sct stars are not stochastically driven and our understanding of driving and damping is poor. For reasons unknown, some stars have clear but incomplete ridges in their \'echelle diagrams: \citet{murphyetal2021a} showed that only half of the radial modes are detected in HD\,139614, and we provide a similar example in Sec.\,\ref{sec:real} using HD\,31901 which is missing half its dipole modes. Hence, we did not design our grid with the intention of modelling the overall excitation behind $\delta$\,Sct p\:modes (for such an analysis, see \citealt{steindletal2021b}), and so we do not have the necessary information to model the frequency of maximum power, $\nu_{\rm max}$. Some studies have suggested that a $\nu_{\rm max}$--$T_{\rm eff}$ relation exists for $\delta$\,Sct stars \citep{barcelofortezaetal2018,barcelofortezaetal2020,hasanzadeh2021}, but TESS observations of 36 Pleiades $\delta$\,Sct stars (of the same age and metallicity) seem to rule out a simple relation \citep{beddingetal2023}. Instead, we turn our attention to further characterising the second scaling relation, which relates $\Delta\nu$ to the square root of the mean stellar density.

\begin{figure}
\begin{center}
\includegraphics[width=0.48\textwidth]{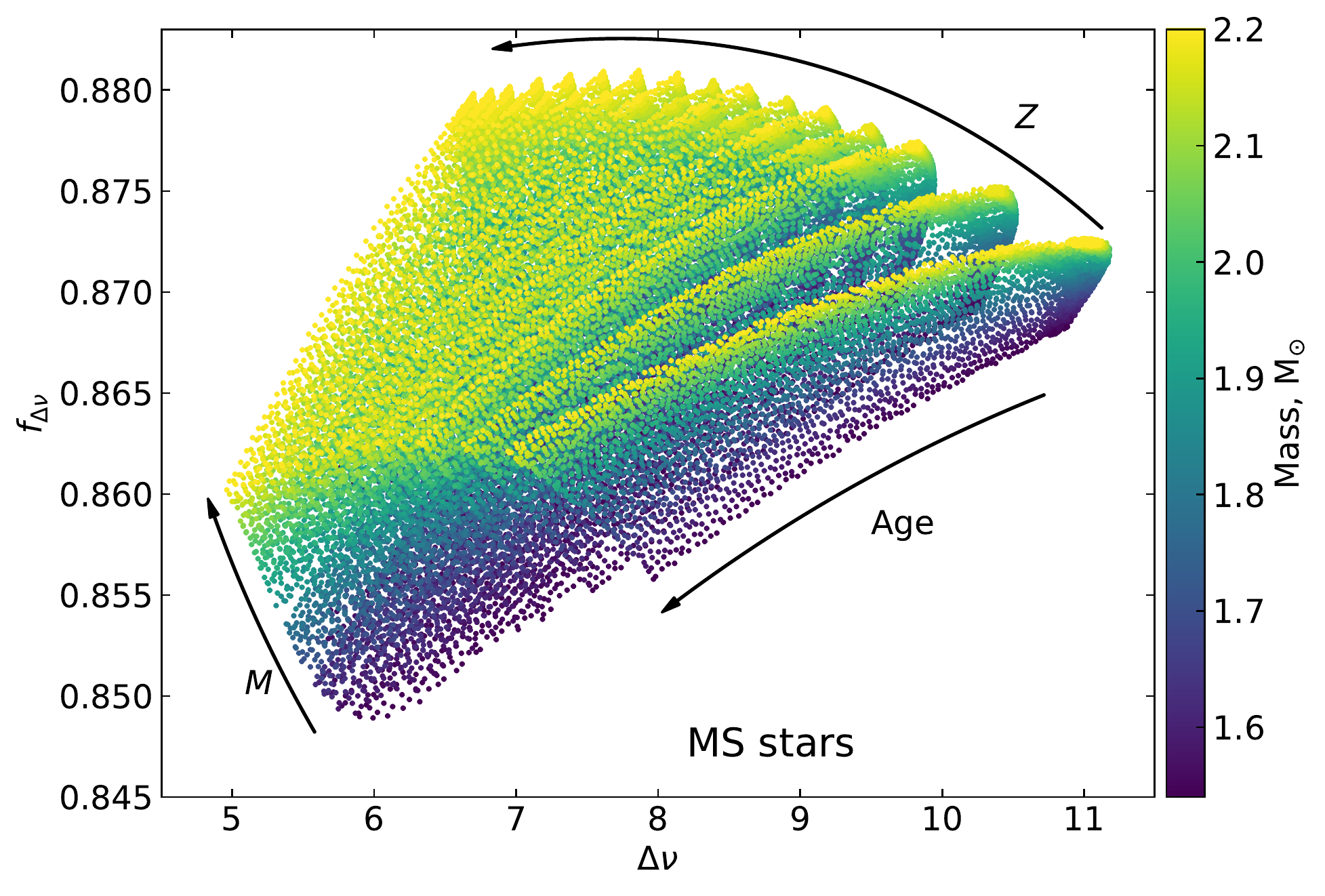}
\caption{Departure of $\Delta\nu$ from the asteroseismic scaling relation. There is a systematic offset of $\sim$0.865 with small but predictable perturbations according to mass, metallicity, and age. Models shown here have $M\geq1.53$\,M$_{\odot}$ and $Z\geq0.002$.}
\label{fig:fDnu}
\end{center}
\end{figure}


In Fig.\,\ref{fig:fDnu} we plot departure of $\Delta\nu$ from the scaling relation, following eq. 5 of \citet{sharmaetal2016}:
\begin{eqnarray}
f_{\Delta\nu} = \left( \frac{\Delta\nu}{\Delta\nu_{\odot}} \right) \left( \frac{\rho}{\rho_{\odot} }\right)^{-0.5}.
\end{eqnarray}
Unlike solar-like oscillators, which have $f_{\Delta\nu}$ within a few percent of unity (\citealt{whiteetal2011c,hekker2020}; see also \citealt{guggenbergeretal2016,rodriguesetal2017,serenellietal2017,pinsonneaultetal2018}), we determine that the $\delta$\,Sct stars have $f_{\Delta\nu} \sim 0.87$. \citet{suarezetal2014} previously determined in a general sense that for $\delta$\,Sct stars $\Delta\nu/\Delta\nu_{\odot} = 0.776(\rho/\rho_{\odot})^{0.46}$. Here, we show that there is clear dependence of $f_{\Delta\nu}$ on our three independent variables, $M$, $Z$, and age. Hence, not only do our models support the existence of a $\Delta\nu$ scaling relation, but they indicate that departures from it might carry information on specific parameters. Ultimately, now that mode identification has become tractable and modelling has become quick \citep{scuttetal2023}, $\delta$\,Sct modelling might skip the era of scaling relations and proceed directly to frequency modelling.

\subsection{The ZAMS $\Delta\nu$ is insensitive to Mass}
\label{ssec:zams_dnu}

\begin{figure}
\begin{center}
\includegraphics[width=0.48\textwidth]{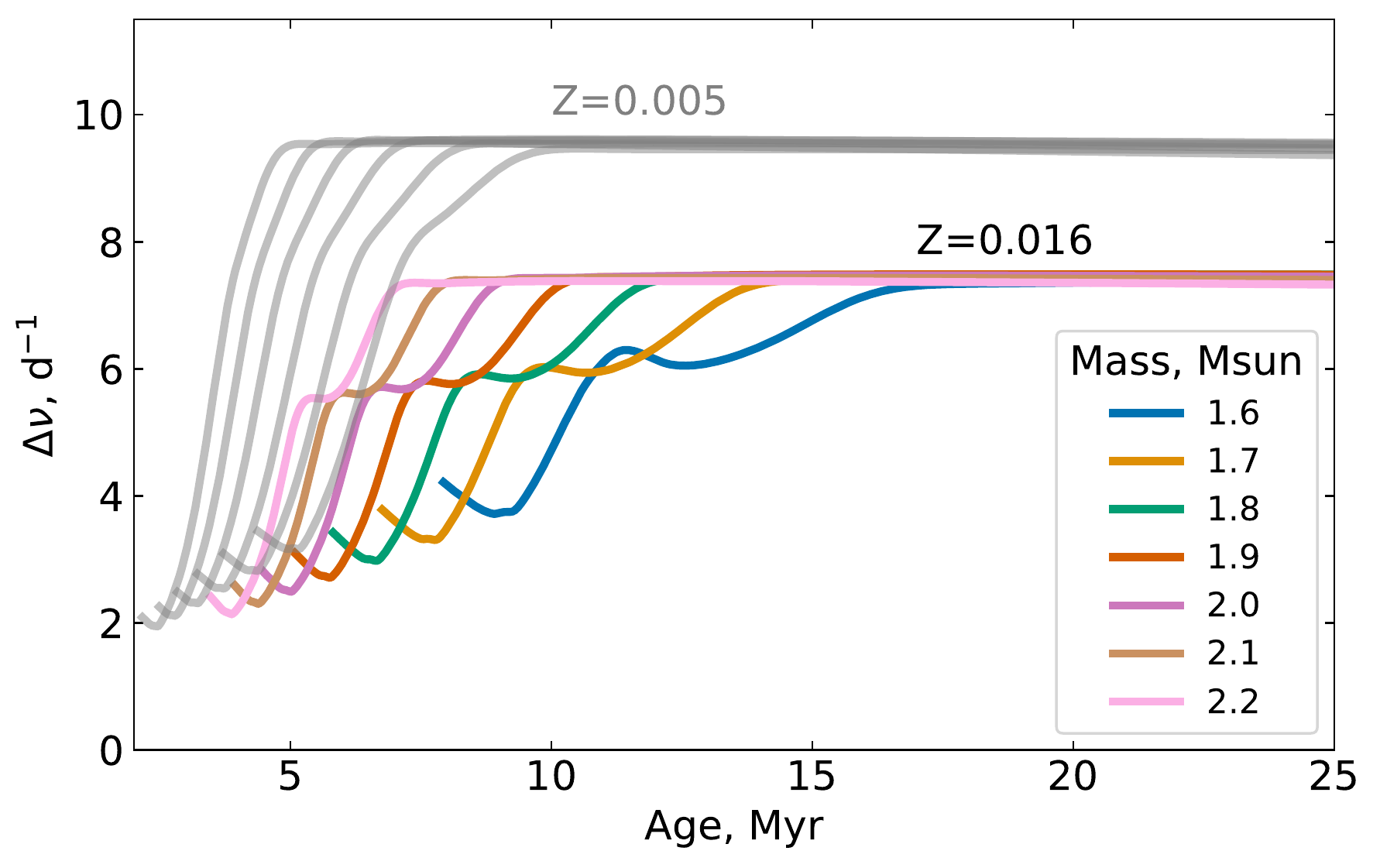}
\caption{The evolution of $\Delta\nu$ as a function of age. Coloured lines represent different masses at Z=0.016; grey lines show the same set of masses at a different metallicity (Z=0.005). The MS value of $\Delta\nu$ is dependent only on metallicity; mass has very little effect. This can also be seen by the near-orthogonality of the mass and metallicity data axes in Figs\,\ref{fig:dnu-eps-ms}\,\&\,\ref{fig:dnu-eps-prems}.}
\label{fig:MS_Dnu}
\end{center}
\end{figure}

\begin{figure}
\begin{center}
\includegraphics[width=0.48\textwidth]{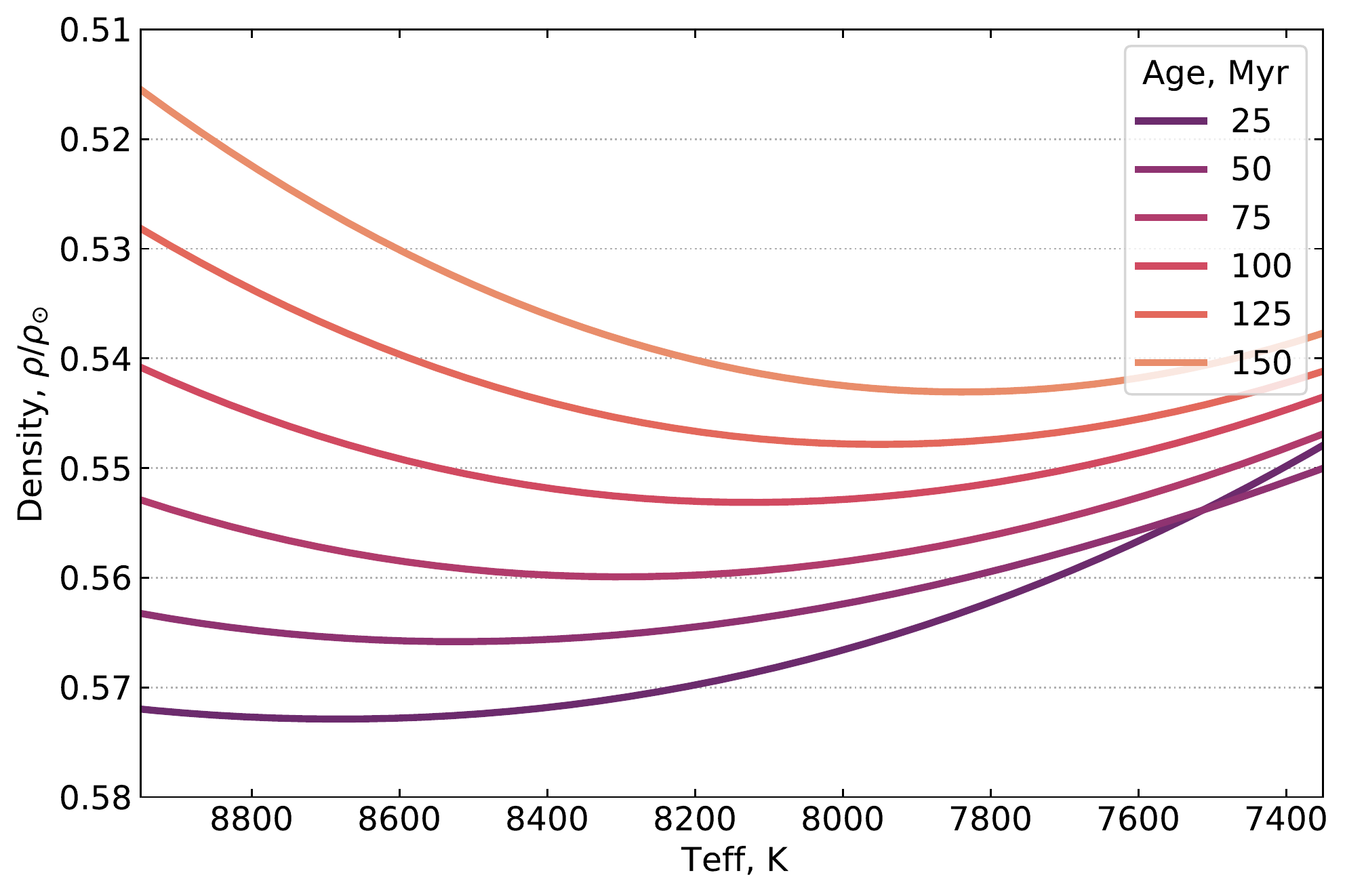}
\caption{Isochrones for the early MS at 25\,Myr spacing, made from tracks of solar metallicity ($Z=0.014$) at masses of 1.38 to 1.70\,M$_{\odot}$. The isochrones are almost flat, having turning points in density over a wide range of temperatures covering the $\delta$\,Sct instability strip. Consecutive isochrones have very similar shapes, despite spanning a wide mass range. Hence, an isochrone for a given metallicity has a nearly constant density as a function of mass and temperature across the instability strip.}
\label{fig:isochrones}
\end{center}
\end{figure}

Our models show that the MS value of $\Delta\nu$ is dependent only on metallicity -- mass has very little effect. Fig.\,\ref{fig:MS_Dnu} shows this mass independence, and also shows that once stars reach the ZAMS, their densities plateau. They do this for around 100\,Myr before they more quickly decrease with age. Because these stars are close to the ZAMS, their isochrones lie parallel to the ZAMS, hence any mass difference has had little differential effect on the stellar evolution. It is interesting that stars of such a wide range of masses arrive on the ZAMS with almost identical densities. Fig.\,\ref{fig:isochrones} shows that isochrones lie parallel to isodensity contours, to a good approximation.

We explored this numerically in more detail. Within the instability strip (at $T_{\rm eff}$ between 7000 and 9000\,K), the density of a 100-Myr isochrone never deviates from its average value of 0.547$\rho_{\odot}$ by more than 1.25\%. Since $\Delta\nu \propto \sqrt{\rho}$, then $\Delta\nu$ should be constant to around 0.6\% (or 0.04\,d$^{-1}$) for a given metallicity. It is remarkable that this occurs despite spanning a mass range of 20\%.\footnote{Models of the period spacings of g-modes in $\gamma$\,Dor stars \citep{mombargetal2019} also appear to be insensitive to mass at the ZAMS to some extent.}
In young clusters such as the Pleiades, this means that rotation is the only factor causing differences in $\Delta\nu$. At rotation velocities of 150\,km\,s$^{-1}$, the difference in density between a static and a rotating model reaches around 8\% \citep{murphyetal2022a}, so $\Delta\nu$ should differ by approx 0.3\,d$^{-1}$ for rapid rotators. The observed spread in $\Delta\nu$ values currently sits at 0.15\,d$^{-1}$ for Pleiades stars with $\Delta\nu$ measurements \citep{murphyetal2022a}.

\subsection{The fundamental radial mode is poorly modelled}

In many multi-periodic stars, the observed frequency of the fundamental radial mode ($f_1$) is poorly modelled, including the well-studied pre-MS star HD\,139614 \citep[][see also Sec.\,\ref{ssec:hd139614}, this work]{murphyetal2021a,steindletal2022}, the otherwise well-modelled star HD\,99506 \citep{scuttetal2023}, and two of the five stars modelled in Sec.\,\ref{sec:real}. This is alarming because traditional methods of identifying modes include the assumption that the strongest mode is radial \citep[e.g.][]{sanchezariasetal2017}, and/or to compare the period ratios of low-frequency p\:modes against those computed with models of radial modes, especially via so-called Petersen diagrams \citep{petersen&christensen-dalsgaard1996,suarezetal2006,netzeletal2021}. The stars mentioned above have models that match well for the entire radial ridge but not for $f_1$.

\begin{figure}
\begin{center}
\includegraphics[width=0.48\textwidth]{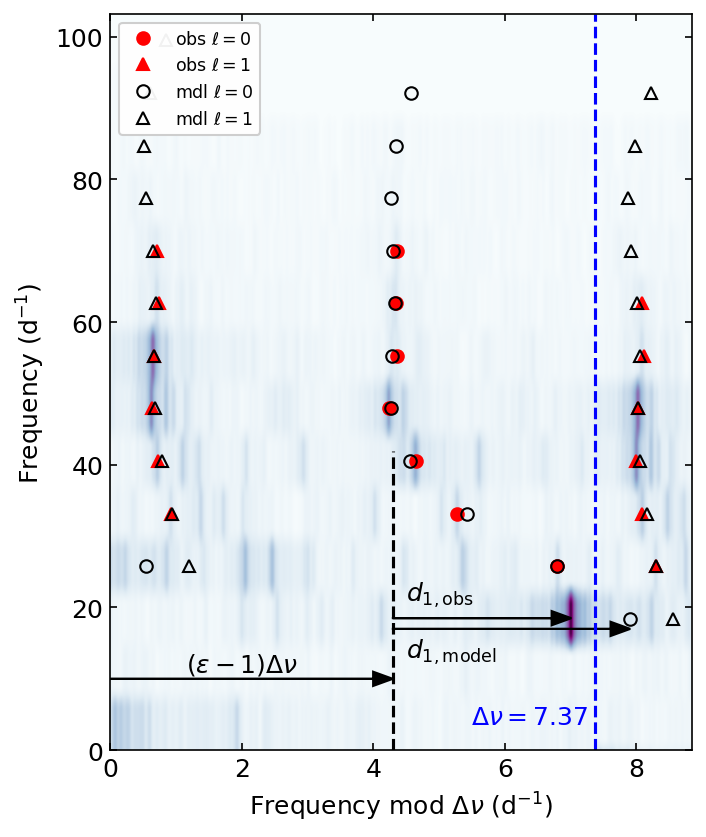}
\caption{
An illustration of the mismatch between models and observations for the fundamental radial mode, using an \'echelle diagram of HD\,20203. The grey-scale shows the observed amplitude spectrum, the red points mark the observed identified modes, and the black symbols are the model frequencies. The vertical dashed blue line is at $\Delta\nu$, and everything to the right of that line is a repetition of the first 20\% of the row above \citep{bedding2012}. The lowest frequency identified mode of each degree is at $n=2$. The highest frequency radial mode present is the $n=8$ mode, hence for this star $\Delta\nu$ and $\epsilon$ are measured for modes of $n=5$ to $8$ (using eq.\,\ref{eq:deltanu}).
We also introduce the $d_1$ parameter, which describes the curvature of the radial ridge at low radial orders. We show $d_1$ for the strong peak at low frequency and for the best-fitting model from the grid. Neither the radial nor the dipole mode at $n=1$ are well modelled.}
\label{fig:d1_echelle}
\end{center}
\end{figure}

Here, we attempt to characterise differences between obvious candidates for $f_1$ (strong peaks at roughly the expected frequency) and their equivalent model frequencies. We use HD\,20203 to illustrate a particularly egregious mismatch. This star has 14 modes that match models very well, but the strong peak near $f_1$ is a poor fit (Fig.\,\ref{fig:d1_echelle}). Although we focus our efforts on the radial mode for simplicity, we have also noticed that matches for the lowest-order dipole mode are often poor, such as in HD\,20203.

\begin{figure}
\begin{center}
\includegraphics[width=0.48\textwidth]{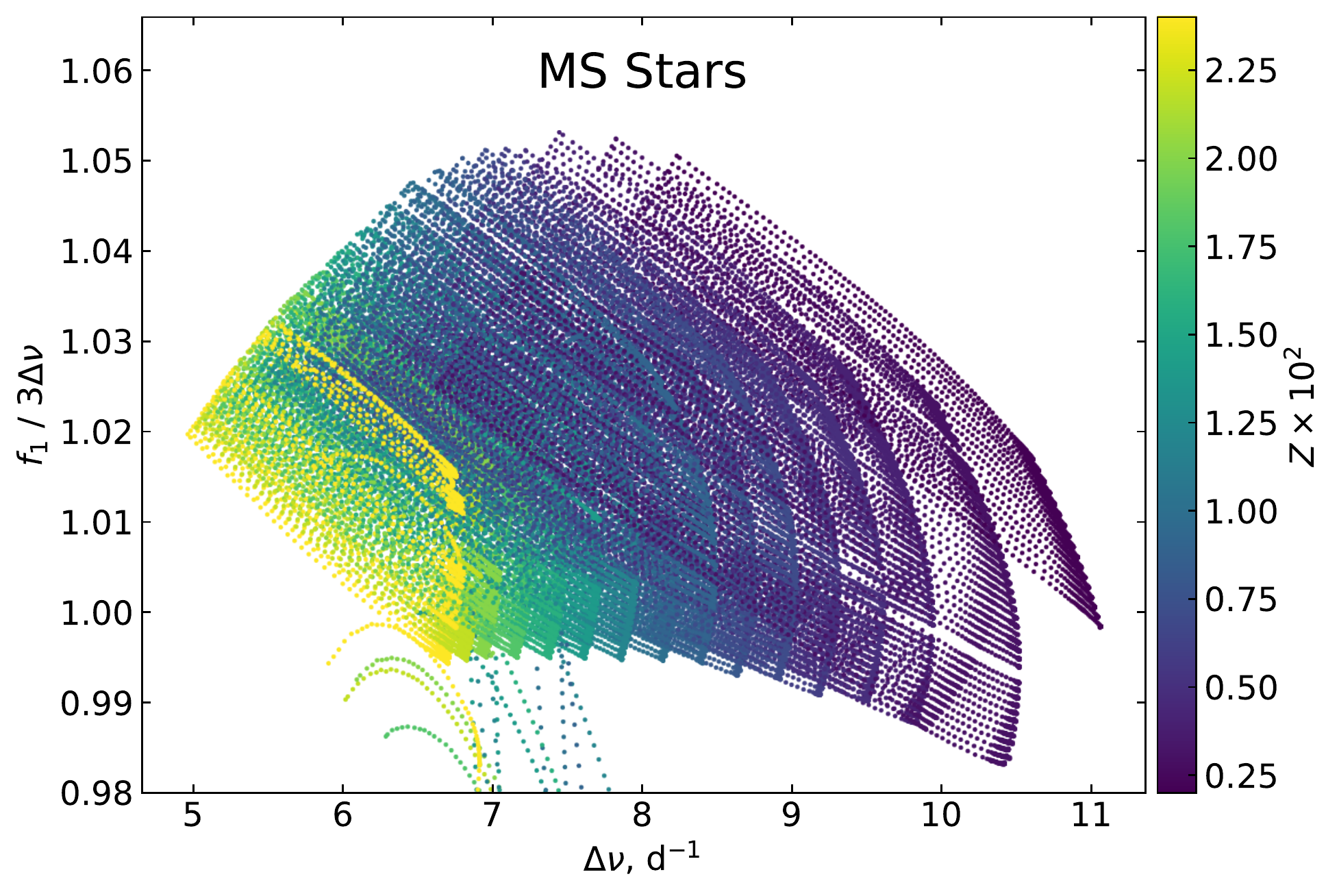}
\caption{The ratio of the fundamental radial mode frequency, $f_1$, to three times the large frequency separation, $3\Delta\nu$, for MS models in the grid. The smattering of points at $\Delta\nu\sim7$ with $f_1/\Delta\nu<0.995$ are mislabelled pre-MS models.}
\label{fig:3dnu}
\end{center}
\end{figure}

Other methods for identifying $f_1$ also exist. In \citet[][their extended data figure 2]{beddingetal2020}, $f_1$ was observed to lie at a frequency of approximately 3$\Delta\nu$. Our model grid shows this to be a good approximation (Fig.\,\ref{fig:3dnu}). Generally, $f_1$ exceeds 3$\Delta\nu$ by a few percent, hence one should expect the radial mode to lie on the LHS of row four of the \'echelle (or 0--1\,d$^{-1}$ to the right of the dashed line of row three in our phase-wrapped \'echelles). It is rare for $f_1$ to lie at frequencies lower than 3$\Delta\nu$ in our models, especially at higher metallicity. \citet{murphyetal2020b} used the 3$\Delta\nu$ approximation and the rest of the radial ridge in \'echelle diagrams to identify the radial mode in 11 pulsating TESS $\lambda$\,Boo stars and used either Petersen diagrams or the Period--Luminosity relation \citep{mcnamara1997,ziaalietal2019,baracetal2022} to identify the fundamental mode in 17 others. It is noteworthy that while \citet{murphyetal2020b} sometimes identified the fundamental mode using two techniques for the same star, Petersen diagrams (i.e. period ratios) and \'echelle diagrams were not successfully applied together.

\begin{figure}
\begin{center}
\includegraphics[width=0.48\textwidth]{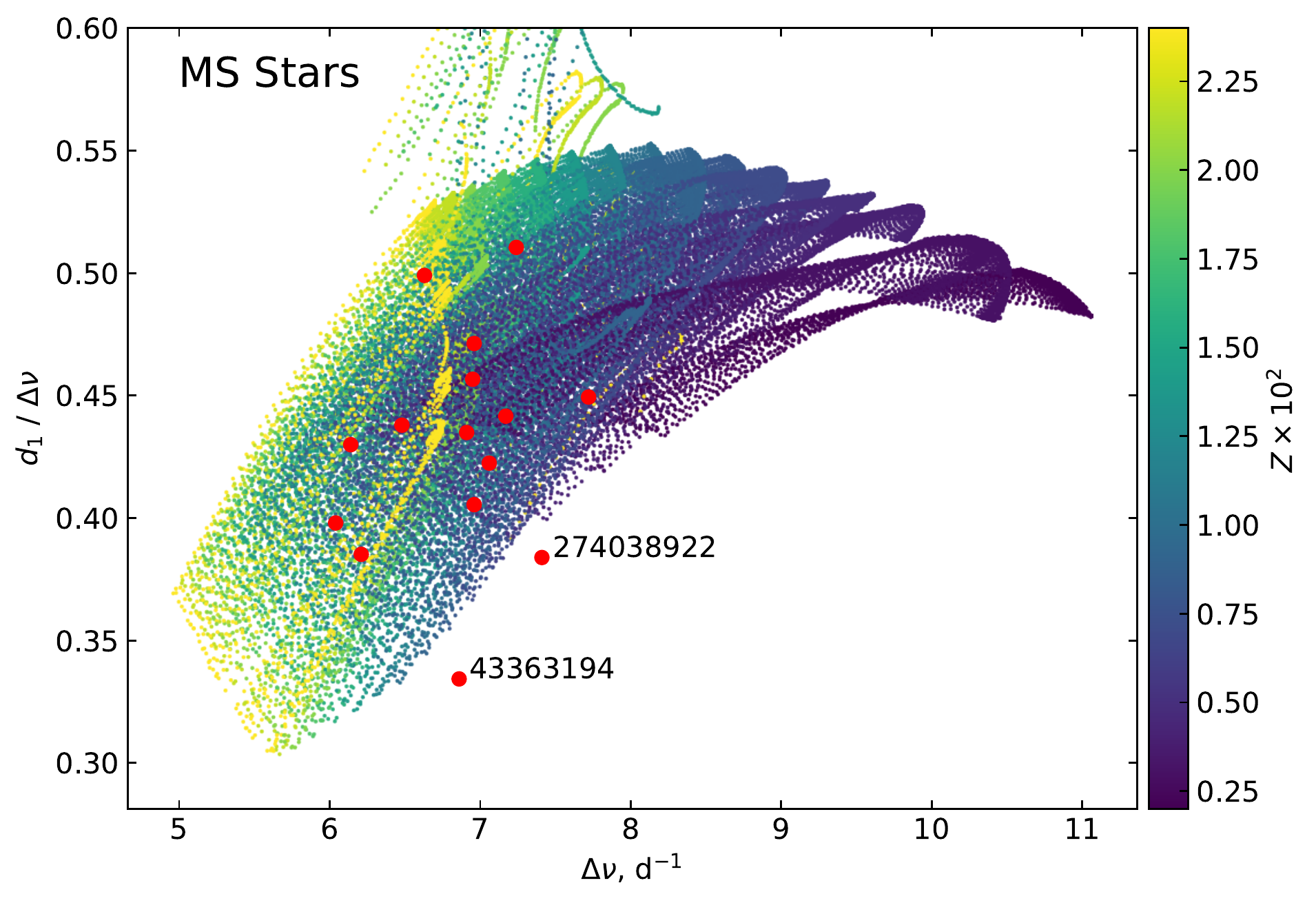}
\includegraphics[width=0.48\textwidth]{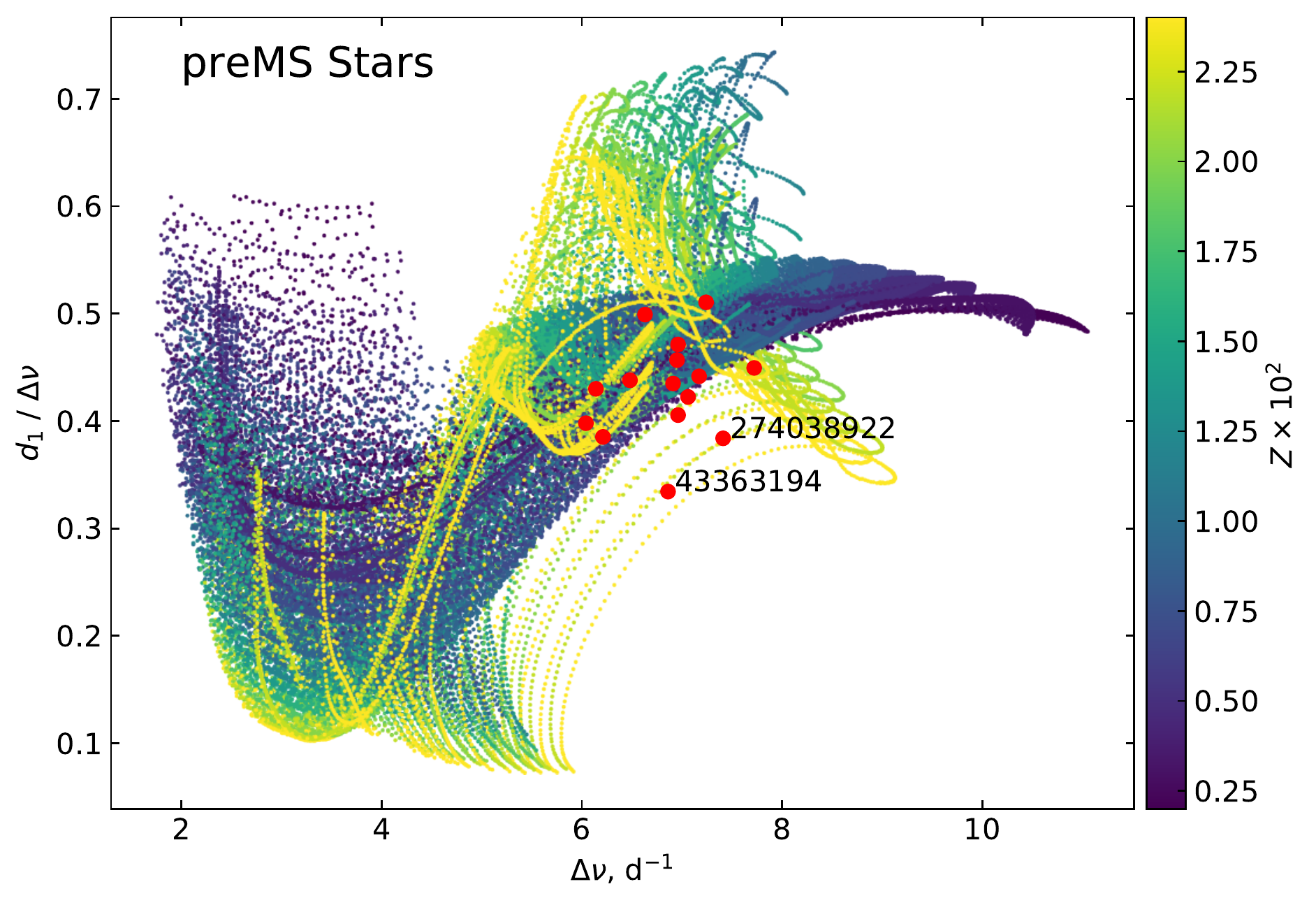}
\caption{The $d_1$ parameter for MS and pre-MS stars. Note that the panels have different axis ranges but a common colour bar. Red circles show the 15 stars from Table\:\ref{tab:nature}. Outliers are TIC\,274038922 = HD\,20203, and TIC\,43363194 = HD\,3622.}
\label{fig:d1}
\end{center}
\end{figure}

To further investigate the aforementioned common mismatch, we define a new quantity, $d_1$, describing the departure of the fundamental radial mode, $f_1$, from the radial ridge in an \'echelle diagram. As shown in Fig.\,\ref{fig:d1_echelle}, it measures the curvature of the radial ridge at low radial orders. The $x$-location of the radial mode ridge is already established via the asteroseismic parameters $\epsilon$ and $\Delta\nu$ (see Sec.\,\ref{ssec:Dnu}), hence
\begin{eqnarray}
d_1 = f_1 - (\epsilon+1)\Delta\nu.
\end{eqnarray}
By defining $d_1$ in this way, it is insensitive to the natural variation of $\epsilon$ between stars, unlike the quantity $f_1/3\Delta\nu$, which is simultaneously $\epsilon$ and $\Delta\nu$ dependent. In Fig.\,\ref{fig:d1} we plot $d_1$ in the dimensionless form $d_1/\Delta\nu$ for models in our grid, separated into pre-MS and MS stages of evolution. To this figure, we have again added the 15 stars from Table\,\ref{tab:nature}, and we note that their evolutionary states (pre-MS vs. MS) are unknown.

As expected, HD\,20203 is one of the outliers in Fig.\,\ref{fig:d1}, indicating that the strong peak at 21.75\,d$^{-1}$ is incompatible with the fundamental radial mode in our models. We do not expect this to be fixed by the inclusion of rotation, since a change large enough to shift $f_1$ by $\sim$1\,d$^{-1}$ will also change other mode frequencies substantially. Physics that affects only the longer-wavelength (deeper-penetrating) p\:modes and their period ratios most likely concerns the near-core region, hence the degree of overshooting from the convective core might be responsible. We will investigate this in future work.

In summary, the $d_1$ parameter is a useful check on whether the $f_1$ has been correctly identified. Agreement with models is a necessary (but insufficient) condition for correct identification of $f_1$. The $d_1$ parameter may also help to diagnose any mismatch of $f_1$, and might be sensitive to core overshooting.


\section{Application to real stars}
\label{sec:real}

Preliminary versions of the grid have already been used to model several stars, including
the superjovian-exoplanet host HIP\,99770 \citep{currieetal2023a}, 
three stars in the new Cepheus Far North association (TIC\,373018187, TIC376872090, and  TIC\,429019921; \citealt{kerretal2022a}), 
five members of the Pleiades star cluster \citep{murphyetal2022a}, and
HD\,21434 in the Fornax--Horologium association \citep{kerretal2022b}. Here we apply the grid to six additional stars to demonstrate its utility. We use the neural network described in \citet{scuttetal2023} to perform the Bayesian inference and to provide quantitative uncertainties on mass, age, and metallicity. 

For all stars modelled here, the likelihood of the observations, given the input model parameters, $\theta$, is given by eq. 4 of \citet{scuttetal2023}
\begin{equation}
    \log{\mathcal{L}\left(D|\theta\right)} = \log{\mathcal{L}(D_{\mathrm{S}}|\theta)} + \log{\mathcal{L}(D_{\mathrm{C}}|\theta)},
\end{equation}
which we separate into the seismic contribution, $D_\mathrm{S}$, and classical (non-seismic) contributions $D_\mathrm{C}$, comprising of the effective temperature $T_{\rm eff}$ and luminosity $L$. The contribution to the likelihood of the mode frequencies is given by
\begin{equation}
   \log{\mathcal{L}(D_{\mathrm{S}}|\theta)} = \sum_i \log \mathcal{N}\left( \nu^{\mathrm{obs}}_i, \sqrt{\sigma_{\nu^{\mathrm{obs}}_i}^2 + \sigma_{\nu^{\mathrm{NN}}_i}^2} \right),
\end{equation}
and that of the classical observables is given by
\begin{align}
   \log{\mathcal{L}(D_{\mathrm{C}}|\theta)} = &\log\mathcal{N}\left(\log{L}^{\mathrm{obs}}, \sqrt{\sigma_{L^{\mathrm{obs}}}^2 + \sigma_{L^{\mathrm{NN}}}^2}\right) \,+ \nonumber \\
   &\log \mathcal{N}\left(T_{\rm eff}^{\mathrm{obs}}, \sqrt{\sigma_{T_{\rm eff} ^{\mathrm{obs}}}^2 + \sigma_{T_{\rm eff}^{\mathrm{NN}}}^2}\right)
\end{align}
(\citeauthor{scuttetal2023} eqs 5\,\&\,6). The intrinsic uncertainty arising from the neural network representation of the data, $\sigma_{\rm NN}$, is captured in each case. Further details can be found in \citet{scuttetal2023}.

\subsection{A worked example: HD\,139614}
\label{ssec:hd139614}

We start with the protoplanetary disk host HD\,139614, which has been modelled by both \citet{murphyetal2021a} and \citet{steindletal2022}. The mode IDs were common across both studies, though \citeauthor{steindletal2022} also labelled the mode at 20.599\,d$^{-1}$ as the fundamental radial mode, where \citeauthor{murphyetal2021a} had declined to use it because it didn't match their models. We note here that $d_1/\Delta\nu = 0.442$ for that mode, which is compatible with both the pre-MS and MS distribution in Fig.\,\ref{fig:d1}, but nonetheless we find it incompatible with the other (well-matched) ids. Another main difference between these two studies was the helium abundance. \citeauthor{murphyetal2021a} fixed the helium abundance to $Y=0.29$, whereas \citeauthor{steindletal2022} allowed a very broad range from 0.216 to 0.282 in order to achieve low helium abundances in metal-rich models via their helium enrichment equation (they used $dY/dZ=2.0$ rather than our $1.4$ in their equivalent of our Eq.\,\ref{eq:dydz}), a consequence of which is that their helium abundance may be lower than the primordial abundance from big-bang nucleosynthesis in some cases. For HD\,139614, having $Z\approx0.010$, our helium abundance in this work is $Y=0.2741$. Finally, \citeauthor{steindletal2022} used three modelling approaches with regards to the classical observables: one where only a 1$\sigma$ box was taken around the observed values (i.e. the approach used by \citealt{murphyetal2021a}); another where a 3$\sigma$ box was used; and a final approach where the $\chi^2$ of the classical observables was added to the asteroseismic $\chi^2$ (their eq. 7). We agree with \citeauthor{steindletal2022} that the 1$\sigma$ approach is too narrow. Our approach here is similar to their $\chi^2$ addition formula, except that we do not need to specify $\chi^2$ thresholds based on $p$\:values.

\begin{figure}
\begin{center}
\includegraphics[width=0.48\textwidth]{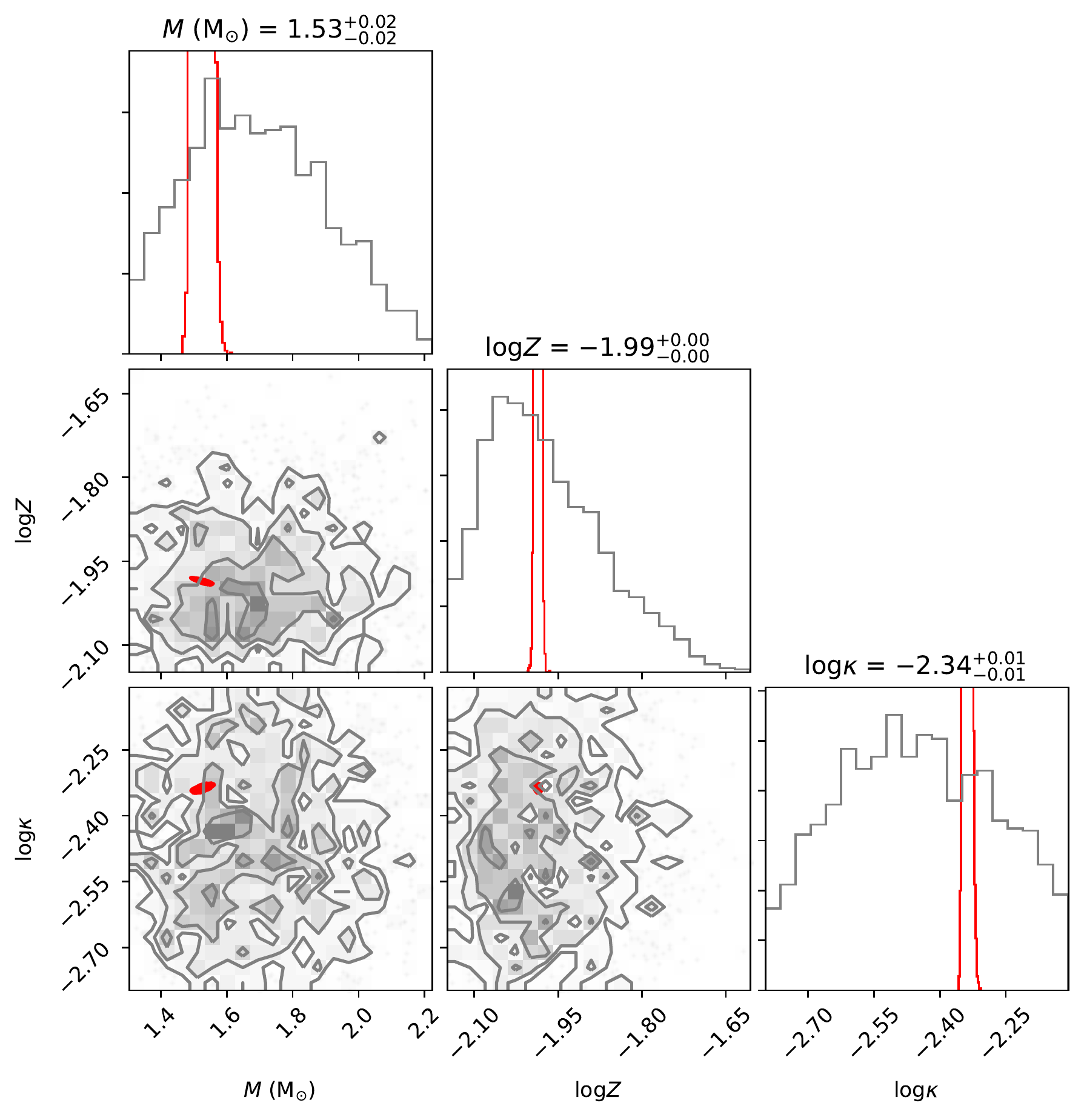}
\caption{The sampled priors applied when modelling HD\,139614 (grey contours and histograms). $\chron$ is a scaled age parameter similar to a fractional age, with the histogram here spanning $\sim$3--15\,Myr. Also shown are the posterior distributions for HD\,139614 (red), which are contained well within the bounds of those priors. The parameters are described in Sec.\,\ref{ssec:hd139614}.}
\label{fig:hd139614_priors}
\end{center}
\end{figure}

Compared to other stars we model here, we treated HD\,139614 as a special case because it has a measured metallicity in the literature and it has already been modelled asteroseismically. As \citet{murphyetal2021a} mentioned, the literature [Fe/H] value ($-0.57\pm0.13$, \citealt{folsometal2012}) is somewhat biased by a chemical peculiarity of the $\lambda$\,Boo type \citep{kamaetal2015}, and does not correspond to the best-fitting asteroseismic solution. Nonetheless, we did not wish to exclude asteroseismic solutions that would match the lower spectroscopic [Fe/H] value. We therefore chose for our beta-distribution prior on $\log Z$: $\beta^{2}_{6}(-2.15, 0.7)$. We plot all of our priors in Fig.\,\ref{fig:hd139614_priors}. For the age prior, rather than following \citet{murphyetal2021a} and imposing a flat prior with a maximum age of 30\,Myr, we chose a prior that strongly preferences ages compatible with its membership in Upper Centaurus--Lupus. UCL has a median cluster age of $16\pm2$\,Myr \citep{pecaut&mamajek2016}, with a 1$\sigma$ age spread of 7\,Myr \citep{mamajeketal2002,preibisch&mamajek2008,pecaut&mamajek2016}, and HD\,139614 has an asteroseismically modelled age of 10--12\,Myr \citep{murphyetal2021a,steindletal2022}. Specifically, we adopted $\log \chron$: $\beta^{2}_{1.2}(-3, -0.3)$ for this star (where $\chron$ is a scaled age parameter similar to a fractional age; see \citealt{scuttetal2023} for details). For the mass prior we used the beta distribution in \citeauthor{scuttetal2023}, namely $M [{\rm M}_{\odot}]$: $\beta^{2}_{3}(1.3, 2.3)$, which allows any mass between $\sim$1.3 and 2.2\,M$_{\odot}$. The best-fitting solution is contained well within these priors (Fig.\,\ref{fig:hd139614_priors}).

We found a best-fitting mass of $1.53\pm0.02$\,M$_{\odot}$, a metal mass fraction of $Z = 0.0103\pm0.0001$ and an age of $11.87\pm0.33$\,Myr. We emphasize that the uncertainties only represent the random uncertainty, including that inherent within the neural network itself. Systematic uncertainty pertaining to model physics remains unaccounted for because it is not well understood, and a thorough analysis of that uncertainty is urgently needed. Nonetheless, we can see that the resulting age is intermediate between that of \citet{murphyetal2021a} and \citet{steindletal2022} ($10.75\pm0.77$ and $12\pm3$\,Myr, respectively), probably as a result of having an intermediate helium abundance. The application of the neural network has halved the random uncertainties.

\begin{figure*}
\begin{center}
\includegraphics[width=0.98\textwidth]{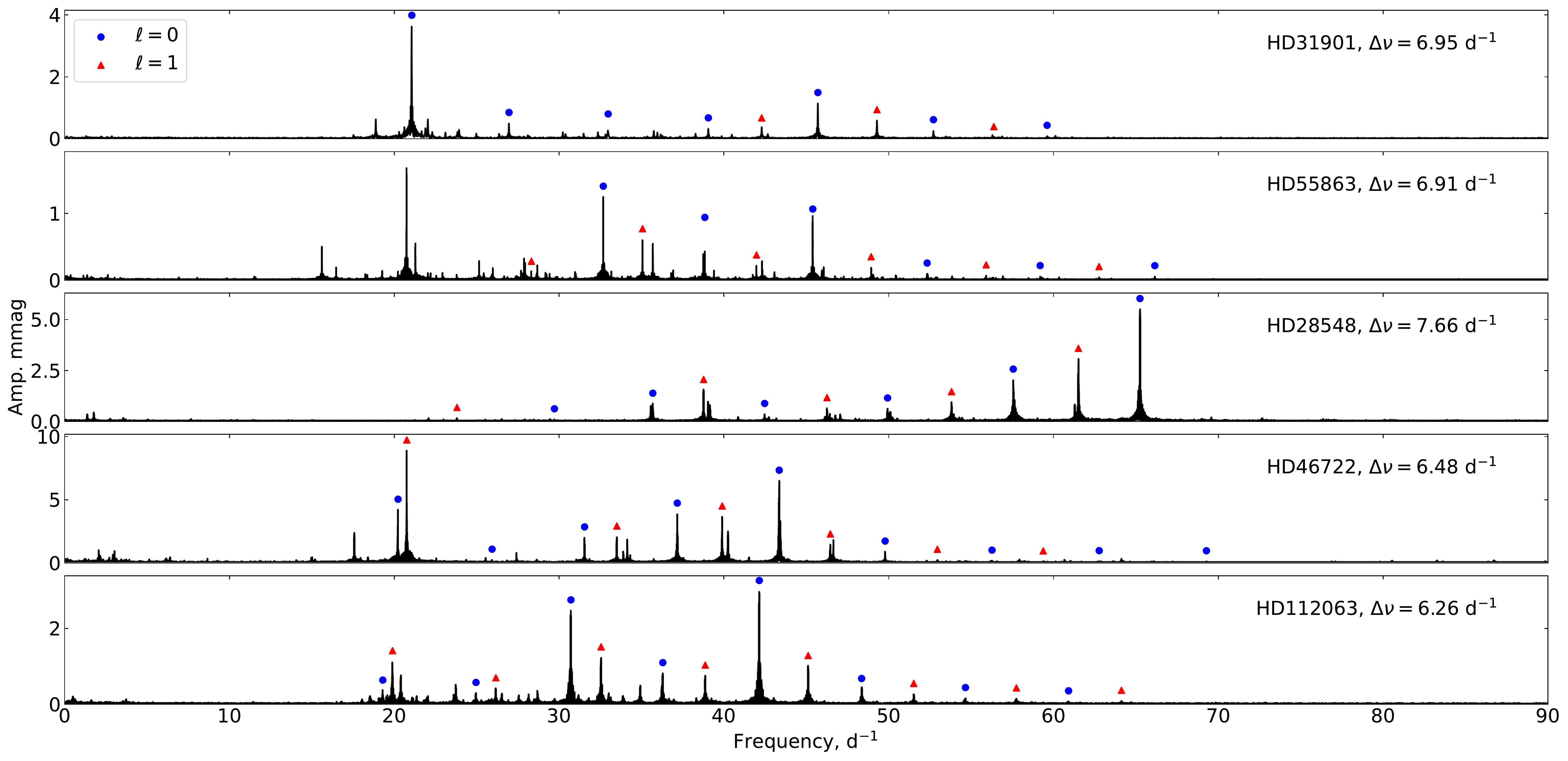}
\caption{Fourier transforms of the TESS light curves of five of the $\delta$\,Sct stars modelled in this work (Sec.\,\ref{ssec:extra_stars}). Blue circles and red triangles label their radial and dipole modes, respectively. These mode IDs are given in Table\:\ref{tab:freqs}.}
\label{fig:FTs}
\end{center}
\end{figure*}

\begin{table*}
\caption{Mode IDs for the five stars analysed here for the first time (Sec.\,\ref{ssec:extra_stars}).}
\label{tab:freqs}
\begin{tabular}{rrrrrrrrrrrrrrrrrrrr}
\toprule
\multicolumn{4}{c}{HD\,31901} & \multicolumn{4}{c}{HD\,55863} & \multicolumn{4}{c}{HD\,28548} & \multicolumn{4}{c}{HD\,46722} & \multicolumn{4}{c}{HD\,112063}\\
\cmidrule(lr){1-4}\cmidrule(lr){5-8}\cmidrule(lr){9-12}\cmidrule(lr){13-16}\cmidrule(lr){17-20}
\multicolumn{1}{c}{Freq.} &  Amp. &  $n$ &  $\ell$ & \multicolumn{1}{c}{Freq.} &  Amp. &  $n$ &  $\ell$ & \multicolumn{1}{c}{Freq.} &  Amp. &  $n$ &  $\ell$ & \multicolumn{1}{c}{Freq.} &  Amp. &  $n$ &  $\ell$ & \multicolumn{1}{c}{Freq.} &  Amp. &  $n$ &  $\ell$ \\
\multicolumn{1}{c}{d$^{-1}$} & $\upmu$mag & & & \multicolumn{1}{c}{d$^{-1}$} & $\upmu$mag & & &\multicolumn{1}{c}{d$^{-1}$} & $\upmu$mag & & & \multicolumn{1}{c}{d$^{-1}$} & $\upmu$mag & & & \multicolumn{1}{c}{d$^{-1}$} & $\upmu$mag & & \\
\midrule
 21.0658 &  3628 &      1 &      0 &  32.6853 &  1240 &      3 &      0 &  29.7205 &    76 &      2 &      0  & 20.2317 &  4152 &    1 &    0 &  19.3056 &   342 &      1 &      0 \\
 26.9651 &   489 &      2 &      0 &  38.8456 &   772 &      4 &      0 &  35.6892 &   839 &      3 &      0  & 25.9366 &   200 &    2 &    0 &  24.9626 &   280 &      2 &      0 \\
 32.9783 &   440 &      3 &      0 &  45.3935 &   898 &      5 &      0 &  42.4728 &   337 &      4 &      0  & 31.5481 &  1965 &    3 &    0 &  30.7183 &  2457 &      3 &      0 \\
 39.0582 &   316 &      4 &      0 &  52.3324 &    89 &      6 &      0 &  49.9313 &   607 &      5 &      0  & 37.1734 &  3845 &    4 &    0 &  36.2954 &   803 &      4 &      0 \\
 45.7033 &  1133 &      5 &      0 &  59.1888 &    51 &      7 &      0 &  57.5579 &  2023 &      6 &      0  & 43.3591 &  6451 &    5 &    0 &  42.1488 &  2968 &      5 &      0 \\
 52.7157 &   256 &      6 &      0 &  66.1454 &    50 &      8 &      0 &  65.2462 &  5483 &      7 &      0  & 49.7844 &   840 &    6 &    0 &  48.3568 &   383 &      6 &      0 \\
 59.6106 &    74 &      7 &      0 &  28.3225 &   116 &      2 &      1 &  23.8091 &   140 &      1 &      1  & 56.2692 &   124 &    7 &    0 &  54.6575 &   145 &      7 &      0 \\
 42.2948 &   315 &      4 &      1 &  35.0640 &   603 &      3 &      1 &  38.7703 &  1515 &      3 &      1  & 62.7715 &    81 &    8 &    0 &  60.9121 &    58 &      8 &      0 \\
 49.2877 &   579 &      5 &      1 &  41.9811 &   210 &      4 &      1 &  46.2566 &   623 &      4 &      1  & 69.2775 &    75 &    9 &    0 &  19.8986 &  1118 &      1 &      1 \\
 56.3789 &    33 &      6 &      1 &  48.9356 &   183 &      5 &      1 &  53.8129 &   917 &      5 &      1  & 20.7644 &  8850 &    1 &    1 &  26.1657 &   403 &      2 &      1 \\
		 &		 &		  &		   &  55.9122 &    62 &      6 &      1 &  61.5114 &  3040 &      6 &      1  & 33.5084 &  2036 &    3 &    1 &  32.5512 &  1218 &      3 &      1 \\
		 &		 &		  &		   &  62.7613 &    36 &      7 &      1 &          &       &        &         & 39.8974 &  3620 &    4 &    1 &  38.8685 &   737 &      4 &      1 \\
         &       &        &        &          &       &        &        &          &       &        &         & 46.4600 &  1410 &    5 &    1 &  45.1138 &   989 &      5 &      1 \\
         &       &        &        &          &       &        &        &          &       &        &         & 52.9526 &   189 &    6 &    1 &  51.5212 &   255 &      6 &      1 \\
         &       &        &        &          &       &        &        &          &       &        &         & 59.3739 &    60 &    7 &    1 &  57.7478 &   135 &      7 &      1 \\
         &       &        &        &          &       &        &        &          &       &        &         & 		&		&	   &	  &  64.1219 &    72 &      8 &      1 \\
\bottomrule
\end{tabular}

\end{table*}

\subsection{Further examples}
\label{ssec:extra_stars}

We also re-examined three stars from \citet{beddingetal2020} with measured $v\sin i$: the slow rotator HD\,31901 ($33\pm4$\,km\,s$^{-1}$), the moderate rotator HD\,55863 ($v\sin i = 99\pm5$\,km\,s$^{-1}$), and the rapid rotator HD\,28548 ($v\sin i = 200\pm50$\,km\,s$^{-1}$). The latter demonstrates that the non-rotating grid can still aid mode IDs for rapid rotators. We also examined two stars whose rotation rates have not been measured: HD\,46722, also from the \citet{beddingetal2020} sample, which has an exceptionally long radial ridge in its \'echelle ($n=1$--9), and HD\,112063, with 8 dipole modes and 8 radial modes (this work). The Fourier transforms of TESS lightcurves of these five stars are shown in Fig.\,\ref{fig:FTs} with their modes labelled. Those labelled mode frequencies are given in Table\:\ref{tab:freqs}.

We analysed all of these stars with the default priors on mass, age, and metallicity from \citet{scuttetal2023}. We used Gaussian priors on temperature and luminosity, specified in Table\:\ref{tab:params}, and we report the posterior probability estimates for their stellar parameters in Table\:\ref{tab:params}. Plots of these estimates are shown in Appendix~\ref{app:stars} (Figs\,\ref{fig:corners1}--\ref{fig:corners3}). We found that three stars (HD\,31901, HD\,55863, HD\,28548) have young MS ages, whereas HD\,46722 and HD\,112063 are in their pre-MS stage. 

{\renewcommand{\arraystretch}{1.25} 
\begin{table*}
\caption{Priors and outputs on stellar parameters.}
\begin{tabular}{r r@{ $\pm$ }l r@{ $\pm$ }l c r@{ $\pm$ }l r@{ $\pm$ }l r@{~}l}
\toprule
 & \multicolumn{4}{c}{Input} & &	 \multicolumn{6}{c}{Output} \\
\cmidrule(lr){2-5}\cmidrule(lr){7-12}
\multicolumn{1}{c}{Star} & \multicolumn{2}{c}{$T_{\rm eff}$ $\pm$ err} & \multicolumn{2}{c}{$L$ $\pm$ err} &	$T_{\rm eff}$ \& $L$ source & \multicolumn{2}{c}{$M$ $\pm$ err} & \multicolumn{2}{c}{$Z$ $\pm$ err} & \multicolumn{2}{c}{Age $\pm$ err} \\
& \multicolumn{2}{c}{K} & \multicolumn{2}{c}{$L_{\odot}$} & & \multicolumn{2}{c}{$M_{\odot}$} & \multicolumn{2}{c}{}  & \multicolumn{2}{c}{Myr} \\
\midrule
HD\,31901	&$	7770	$&$	250	$&$	7.74	$&$	0.39	$&	\citet{beddingetal2020}	&$	1.71	$&$	0.02	$&$	0.0211	$&$	0.0001	$&$	45.04	$&$	^{+9.48}_{-9.01}	$\\
HD\,55863	&$	7650	$&$	250	$&$	7.80	$&$	0.38	$&	\citet{beddingetal2020}	&$	1.75	$&$	0.02	$&$	0.0217	$&$	0.0002	$&$	37.73	$&$	^{+12.51}_{-8.59}	$\\
HD\,28548	&$	8510	$&$	250	$&$	10.82	$&$	0.55	$&	\citet{beddingetal2020}	&$	1.84	$&$	0.02	$&$	0.0144	$&$	0.0001	$&$	15.10	$&$	^{+2.67}_{-2.67}	$\\
HD\,46722	&$	7810	$&$	250	$&$	8.28	$&$	0.40	$&	\citet{beddingetal2020}	&$	1.51	$&$	0.01	$&$	0.0076	$&$	0.0003	$&$	9.04	$&$	^{+0.23}_{-0.25}	$\\
HD\,112063	&$	7517	$&$	250	$&$	6.98	$&$	0.30	$&	\citet{stassunetal2019}	&$	1.61	$&$	0.01	$&$	0.0189	$&$	0.0002	$&$	16.00	$&$	^{+0.33}_{-0.30}	$\\
\bottomrule
\end{tabular}
\label{tab:params}
\end{table*}
{\renewcommand{\arraystretch}{1.0} 

\subsubsection{HD\,31901}
HD\,31901 was modelled superficially by \citet{beddingetal2020}, who noted it as a member of the recently discovered Pisces--Eridanus stream \citet{meingastetal2019}. This association has been estimated to be about 120\,Myr old by gyrochronology \citep{curtisetal2019} and this age has been supported by asteroseismology of HD\,31901 \citep{beddingetal2020}. In a thorough spectroscopic analysis, \citet{hawkinsetal2020} measured a metallicity of [Fe/H]$= -0.03 \pm 0.07$, which also supported an age of $\sim$120\,Myr, while the best-fitting kinematic age is slightly older at $\sim$135\,Myr \citep{roser&schilbach2020}.

Of the five stars we study in detail here, HD\,31901 has the fewest identified modes, though it has much untapped potential via a prograde $\ell=1$ ridge. Naturally, we are unable to model prograde modes using non-rotating models. When analysed using the standard priors, the best-fitting metallicity was $Z = 0.0211$, somewhat larger than expected from the literature values above. We re-ran our analysis, attempting to model the star with a tight solar metallicity prior $Z \approx 0.0144$, but we were unable to obtain a good fit. We note that the posteriors indicated an inverse correlation between metallicity and age, which suggests an age of around 120\,Myr at $Z\sim0.019$. Hence, asteroseismology strongly supports a young MS age, rather than the 1-Gyr age suggested for the Pisces--Eridanus stream by \citet{meingastetal2019}.

\subsubsection{HD\,55863}
This star was one of the examples with mode IDs in \citet[][see their extended data figure 1]{beddingetal2020}, and also appears to have a prograde dipole ridge. We modelled twelve modes, consisting of six consecutive radial orders of the radial and dipole ridges. HD\,55683 has a moderate $v\sin i$ and might make a simple case for study with models that include rotation.

\subsubsection{HD\,28548}
This is a $\lambda$\,Boo star with a large infrared excesses in the WISE W3 (19$\sigma$) and W4 (15$\sigma$) bands \citep{grayetal2017}.
\citet{beddingetal2020} matched this star to a model of mass 1.59\,M$_{\odot}$ and age 270\,Myr. Our model is more massive and younger (see Table\:\ref{tab:params}). The temperature determined from Str\"omgren photometry by \citet[][8490$\pm$170\,K]{murphyetal2020b} is very similar to the one we used here, but their luminosity at 10.04\,L$_{\odot}$ is somewhat smaller than the one we used from \citet{beddingetal2020}. We report values based on the \citet{beddingetal2020} inputs. We repeated our analysis of this star using the parameters from \citet{murphyetal2020b} and found that the resulting metallicity and age were the same within $1\sigma$, but the lower luminosity resulted in a lower mass by 2$\sigma$. The bulk (asteroseismic) metallicity we infer for HD\,28548 is solar, which suggests that the spectroscopic metal-line weakness, also reflected in the Str\"omgren photometry, is only skin-deep. This marks the second demonstration (following HD\,139614; \citealt{murphyetal2021a}) that the $\lambda$\,Boo phenomenon is confined to the stellar surface, as suspected from ensemble studies \citep{paunzenetal2015,murphyetal2020b}.

\subsubsection{HD\,46722}
This is a $\lambda$\,Boo star with an observed infrared excess \citep{grayetal2017}.
\citet{murphyetal2020b} calculated stellar properties from isochrones, using a luminosity derived from Gaia parallaxes and temperatures derived from Str\"omgren photometry. This implied a solar metallicity, [Fe/H]$=-0.016^{+0.12}_{-0.14}$. The metallicity we derive here ($Z=0.0076\pm0.0003$) is the lowest in our sample and corresponds to [Fe/H]$=-0.26$, using \citet{asplundetal2009} solar abundances. The fact that the bulk metallicity is lower than the surface metallicity is inconsistent with the $\lambda$\,Boo classification, perhaps representing a limitation in the Str\"omgren method for a dusty star, or reflecting the unmodelled stellar rotation. A further possibility is that the observed luminosity has been strongly affected by extinction, and this in turn is influencing our Bayesian inference on the asteroseismology. We note that this star's Gaia RUWE value is small (0.96) suggesting it is not a binary, hence the luminosity is much more likely to be underestimated (due to dust) than overestimated (due to a companion).

\subsubsection{HD\,112063}
This is also a $\lambda$\,Boo star, although without an infrared excess \citep{grayetal2017}. Contrary to its metal-weak spectrum, we find that the bulk metallicity is slightly above solar. Its long ridges without any missing modes represent the mode complete mode ID for a $\delta$\,Sct star to date (Fig.\,\ref{fig:pred_ech1}).

\subsection{Posterior frequency predictions}
We show posterior predictions for each mode frequency on the \'echelle diagrams in Fig.\,\ref{fig:pred_ech1} as grey symbols. the spread in posterior frequencies gives a good indication of how well the model is constrained. The red symbols on the \'echelles show the observed frequencies that were used as constraints for the neural network, which we expect to lie within the range of predicted frequencies. If not, it suggests that a mode has been misidentified. The $n=2$ radial mode for HD\,31901 would seem to be such a misidentified mode, perhaps because the star does not pulsate in this mode, or perhaps the identification is correct but the mismatch arises from neglecting rotation. These diagrams can also highlight any weaker peaks that lie within the distribution of posterior predicted frequencies, and which might be used in a second iteration of mode identification. An example of such a mode could be the $n=1$ dipole mode in HD\,31901. However, in this work we only performed one iteration, using the most obvious mode identifications, lest we `reinforce' any emerging best-fitting model from early iterations.

\section{Conclusions}

We have presented and made available a grid of stellar models for young $\delta$\,Scuti stars, with mass, metal mass fraction and age as the independent variables. We computed the asteroseismic parameters $\Delta\nu$ and $\epsilon$ for 200\,000 models and examined how they depend on the independent variables (Figs\,\ref{fig:dnu-eps-ms}\,\&\,\ref{fig:dnu-eps-prems}). The near-orthogonality of the mass and metallicity vectors in the $\Delta\nu$--$\epsilon$ plane makes these parameters useful when individual frequency modelling is not possible. Specifically, at the ZAMS, $\epsilon$ is determined almost exclusively by mass and $\Delta\nu$ is determined almost exclusively by metallicity. In other words, models of a given metallicity have the same $\Delta\nu$ at the ZAMS regardless of their mass. Some regularities in $\Delta\nu$ and $\epsilon$ are also seen in the pre-MS stage but there are more caveats to beware of. We also determined that $\Delta\nu$ for MS $\delta$\,Sct stars deviates from the $\Delta\nu$ scaling relation by $13$\% (Fig.\,\ref{fig:fDnu}).

We described a tendency for the fundamental radial mode, and sometimes the $n=1$ dipole mode, to be poorly fitted by our models. The radial ridge in an \'echelle diagram can be modelled well in many cases, but the fundamental radial mode is rarely matched. We have introduced a parameter, $d_1$, which measures the curvature at the bottom of the radial ridge, to help diagnose this and to aid mode identification. We computed the distribution of $d_1$ from our models and placed 15 stars from \citet{beddingetal2020} amongst that distribution (Fig.\,\ref{fig:d1}). This indicated that the fundamental mode was misidentified in two of those stars. We also described the utility of the $3\Delta\nu$ approximation for identifying the radial mode (Fig.\,\ref{fig:3dnu}).

We have demonstrated the power of combining the grid with a neural network for parameter inference for $\delta$\,Sct stars. We revisited HD\,139614 and calculated new, more precise stellar parameters from models with a more moderate helium abundance. We performed detailed modelling for three $\delta$\,Sct stars from \citet{beddingetal2020} with measured $v\sin i$, and found them all to be very young MS stars with ages $<50$\,Myr. We also presented two $\delta$\,Sct stars that have very long radial and dipole ridges comprising a total of 15 and 16 modes for HD\,46722 and HD\,112063, respectively. We find both of these stars to be in the pre-MS stage, with random uncertainties under 3\%.

In future work we will incorporate rotation into the models, identify rotationally split modes, and quantify the systematic uncertainties arising from various physical and computational parameters. We have evaluated the uncertainty arising from different nuclear reaction networks, and we found that {\sc mesa}'s {\tt basic.net} is unsuitable for modelling $\delta$\,Sct stars. We have made recommendations for nuclear reaction networks that maintain accuracy whilst minimising computation time.



\section*{Acknowledgements}

This research made use of {\sc Lightkurve}, a Python package for Kepler and TESS data analysis \citep{lightkurvecollaboration2018}. We thank Yaguang Li for helpful discussions, and thank the anonymous referee for eliciting clarifications in the paper. SJM thanks Warrick Ball, Guy Davies, Martin Nielsen, and Owen Scutt for discussions during an extended visit to the University of Birmingham, funded by a University of Birmingham Commonwealth Fellowship. SJM was supported by the Australian Research Council (ARC) through Future Fellowship FT210100485. TRB was also supported by the ARC, through DP210103119 and FL220100117. MJ gratefully acknowledges funding of MATISSE: \textit{Measuring Ages Through Isochrones, Seismology, and Stellar Evolution}, awarded by the European Commission through the Horizon 2020 research and innovation programme.

\section*{Data Availability}

\noindent
We provide the model grid in the supplementary file ${\tt grid\_export.csv}$. The contents of this file are described in Sec.\,\ref{ssec:grid-export}. An additional file, ${\tt grid\_export\_readme.txt}$, explains the meaning of each column in that csv file.
We also make available the {\sc mesa} inlist, ${\tt mesa\_inlist}$, and the {\sc gyre} template file, ${\tt gyre\_template}$, as supplementary files. Light curves for the stars discussed in Sec.\,\ref{sec:real} are available at the Mikulski Archive for Space Telescopes (MAST): \url{https://archive.stsci.edu/}.


\bibliographystyle{mnras}
\interlinepenalty=10000
\bibliography{sjm_bibliography} 



\appendix
\section{Comparison of {\sc mesa} Reaction Networks}
\label{app:nuclear}

There are several nuclear reaction networks to consider in the {\sc mesa} suite. Each network is tailored to different astrophysical environments and conditions, varying in the number of isotopes and reactions included.\footnote{Readers should consult the {\sc mesa} documentation for more information on the implementation of various nuclear reactions and more details of these networks.} This section presents a comparison of five {\sc mesa} reaction networks, specifically \texttt{basic.net}, \texttt{hot\_cno.net}, \texttt{pp\_and\_cno\_extras.net}, \texttt{pp\_extras.net}, and \texttt{pp\_extras+hot\_cno.net}. These networks and their differences are described from the next subsection. 

Our specific recommendation for modelling of $\delta$\,Sct stars of all ages is to add the \texttt{pp\_extras} and \texttt{hot\_cno} networks together: \texttt{basic.net} is too inaccurate and should not be used alone, whereas \texttt{pp\_and\_cno\_extras.net} is computationally expensive and includes many reactions not relevant to $\delta$\,Sct stars.\footnote{In retrospect, we would have preferred to have reached this conclusion before computing our grid.}

\subsection{Methodology}

In our grid we used \texttt{pp\_and\_cno\_extras.net}, which is the most complete of the above networks in that it incorporates the greatest number of isotopes and reactions. However, it is correspondingly computationally expensive. We have therefore benchmarked both accuracy and computational time against this network. We did this for a reference model located roughly in the middle of the grid at $M=1.7$\,M$_{\odot}$ and a roughly solar metallicity ($Z=0.015$). We calculated one evolutionary track with each nuclear network and computed their pulsation frequencies using the same methodology as the main grid. We expected the difference in nuclear reaction networks to result in slightly different rates of evolution, hence different stellar densities as a function of age. Since our key observables were the stellar pulsation frequencies, rather than stellar density, we evaluated the frequency differences at fixed ages by interpolating the {\sc gyre} frequencies along individual evolutionary tracks. The interpolated tracks were sampled every 0.1\,Myr.

We found that the $\ell=0$ and $\ell=1$ modes exhibited the same behaviour, and that modes of higher radial order were more strongly affected by the changes in density. Specifically we found for any given model at a particular age, $\delta f/f$ was a constant. Hence, rather than evaluating the effects on any given mode, we determined the mean fractional frequency shift, $\langle \delta f/f \rangle$, for all modes in that model. We show how $\langle \delta f/f \rangle$ changes with age for the five nuclear reaction networks in Fig.\,\ref{fig:nuclear}. The \texttt{basic}, \texttt{hot\_cno}, and  \texttt{pp\_extras} curves show substantial differences from our benchmark throughout the evolution, amounting to as much as a 5\% difference in pulsation frequencies at some ages. For a hypothetical star consistent with those in our grid, having observed mode frequencies as high as 75\,d$^{-1}$ and $\Delta\nu = 7.5$, this means that systematic frequency errors as large as $\Delta\nu/2=3.75$\,d$^{-1}$ would result from using the \texttt{basic} and \texttt{hot\_cno} networks instead of more complete networks. For MS stars, the errors are approximately one order of magnitude smaller, but they remain at least two orders of magnitude greater than the frequency resolution of the TESS data. Hence, use of \texttt{basic.net} is insufficient for asteroseismology of $\delta$\,Sct stars at any ages.

We now focus on reducing the computation time. Fig.\,\ref{fig:nuclear} shows that our custom network, \texttt{pp\_extras+hot\_cno.net} (described in \ref{pp_hot_cno}), performs as well as the \texttt{pp\_and\_cno\_extras} reaction network and reduces the total computation time by 10\%. This is because the cores of $\delta$\,Sct stars do not reach high enough temperatures to produce {\tt f19} from {\tt o18} (cold CNO-IV) before the {\tt o18} can decay to {\tt n15} in the cold CNO-III cycle, hence the {\tt cno\_extras} calculations are not required. We therefore recommend this particular combination of nuclear networks for modelling $\delta$\,Sct stars.

In the remainder of this section, we describe the various nuclear reaction networks in more detail.

\begin{figure}
    \centering
    \includegraphics[width=0.48\textwidth]{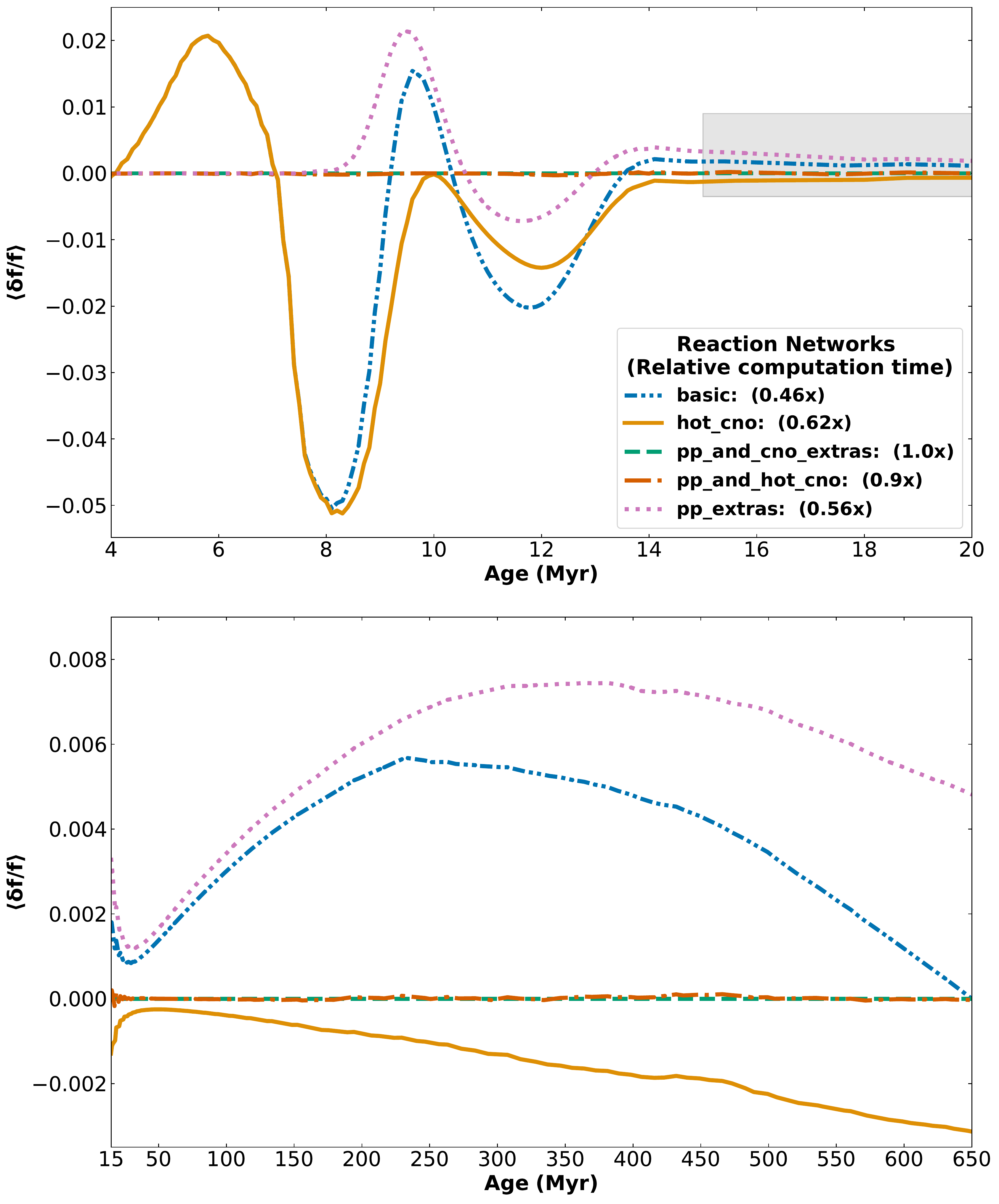}
    \caption{Nuclear reaction networks in {\sc mesa} and their impact on pulsation frequencies. Different curves show the mean fractional change in frequency relative to the baseline of the \texttt{pp\_and\_cno\_extras} reaction network as a function of age. The \texttt{pp\_and\_cno\_extras} curve is therefore at $\langle \delta f/f \rangle$=0 for all ages, and has a relative computational time of unity in the caption. The upper panel focusses on the youngest models (up to 20\,Myr), while the lower panel shows the same from 15 to 650\,Myr. The grey box highlights the overlapping axis range of the second panel with the first. All calculations were performed on tracks with $M = 1.7$\,M$_{\odot}$ and $Z = 0.015$.}
    \label{fig:nuclear}
\end{figure}

\subsection{basic.net}
{\sc MESA}'s \texttt{basic.net} is the simplest reaction network. It assumes that the temperature is low enough to ignore advanced burning and hot carbon--nitrogen--oxygen (CNO) burning. This network includes isotopes \texttt{h1}, \texttt{he3}, \texttt{he4}, \texttt{c12}, \texttt{n14}, \texttt{o16}, \texttt{ne20}, and \texttt{mg24}, covering the proton-proton (pp) chain and the cold CNO-I cycle \citep{krane1988}. Notably, it does not include isotopes \texttt{h2}, \texttt{li7}, \texttt{be7}, and \texttt{b8}, and consequently, it does not account for deuterium burning. The network includes the key reactions in the proton-proton chains, CNO cycles, helium burning, and a set of auxiliary reactions used only for computing rates of other reactions.

\subsection{pp\_extras.net}
The \texttt{pp\_extras} network takes \texttt{basic.net} and extends it by adding the \texttt{pp\_extras}. It introduces isotopes necessary for the inclusion of deuterium burning. It also adds several reactions linked to the pp-chain, and removes certain reactions from the \texttt{basic.net} to accommodate the additional dynamics from the \texttt{pp\_extras}.

\subsection{hot\_cno.net}
The \texttt{hot\_cno} network extends \texttt{basic.net} to include the CNO-II and CNO-III cycles, characterized by high temperatures that permit additional reactions to occur. This may play a crucial role in energy generation in stars more massive than the Sun. Eight isotopes are added: \texttt{c13}, \texttt{n13}, \texttt{n15}, \texttt{o15}, \texttt{o17}, \texttt{o18}, \texttt{f17}, and \texttt{f18}.

To maintain the physical accuracy of the model, a set of reactions primarily related to the CNO cycle from \texttt{basic.net} is removed. The removed reactions represent those processes which are subsumed or replaced by the more sophisticated hot CNO cycle dynamics.

Furthermore, another reaction network, \texttt{cno\_extras.net}, is available for even more extreme conditions, such as those present in massive stars. This network, which includes \texttt{hot\_cno.net} as a foundation, introduces additional isotopes \texttt{o14}, \texttt{f19}, \texttt{ne18}, \texttt{ne19}, and \texttt{mg22} and caters to the hottest reactions (CNO-IV) of the `cold' cycle. However, we find that energy generation rates from \texttt{cno\_extras} and \texttt{hot\_cno} are indistinguishable for our use case. The `cold' CNO-IV cycle is only expected to be important in massive stars, not in $\delta$Sct stars.

\subsection{pp\_and\_cno\_extras.net}
The \texttt{pp\_and\_cno\_extras} network incorporates the extensions provided by both the \texttt{pp\_extras} and \texttt{cno\_extras} networks. It adds the isotopes and reactions corresponding to these two extensions, creating a more comprehensive network that can model both the enhanced pp-chain and the complete CNO cycle over all temperature ranges. This is the most complete set of reactions we consider.

\subsection{pp\_extras+hot\_cno.net}
\label{pp_hot_cno}
The \texttt{pp\_extras\_and\_hot\_cno} network is a custom {\sc mesa} reaction network that is not included as a default option. We specifically constructed it by combining \texttt{basic.net} with both the \texttt{pp\_extras} and \texttt{hot\_cno} extensions. This unique configuration enables the network to incorporate the necessary isotopes and reactions from both extensions, resulting in a comprehensive model that is comparable to the \texttt{pp\_and\_cno\_extras} network, without the unneeded CNO-IV calculations. Thus, \texttt{pp\_and\_hot\_cno.net} offers faster computational performance than \texttt{pp\_and\_cno\_extras.net} whilst maintaining accuracy.

\section{Additional figures for the five newly modelled stars}
\label{app:stars}

In this appendix we provide the corner plots for the five modelled stars (refer to Sec.\,\ref{sec:real}) and the corresponding \'echelle diagrams with posterior-predicted mode frequencies.

\begin{figure}
\begin{center}
\begin{overpic}[abs,unit=1mm,scale=0.48]{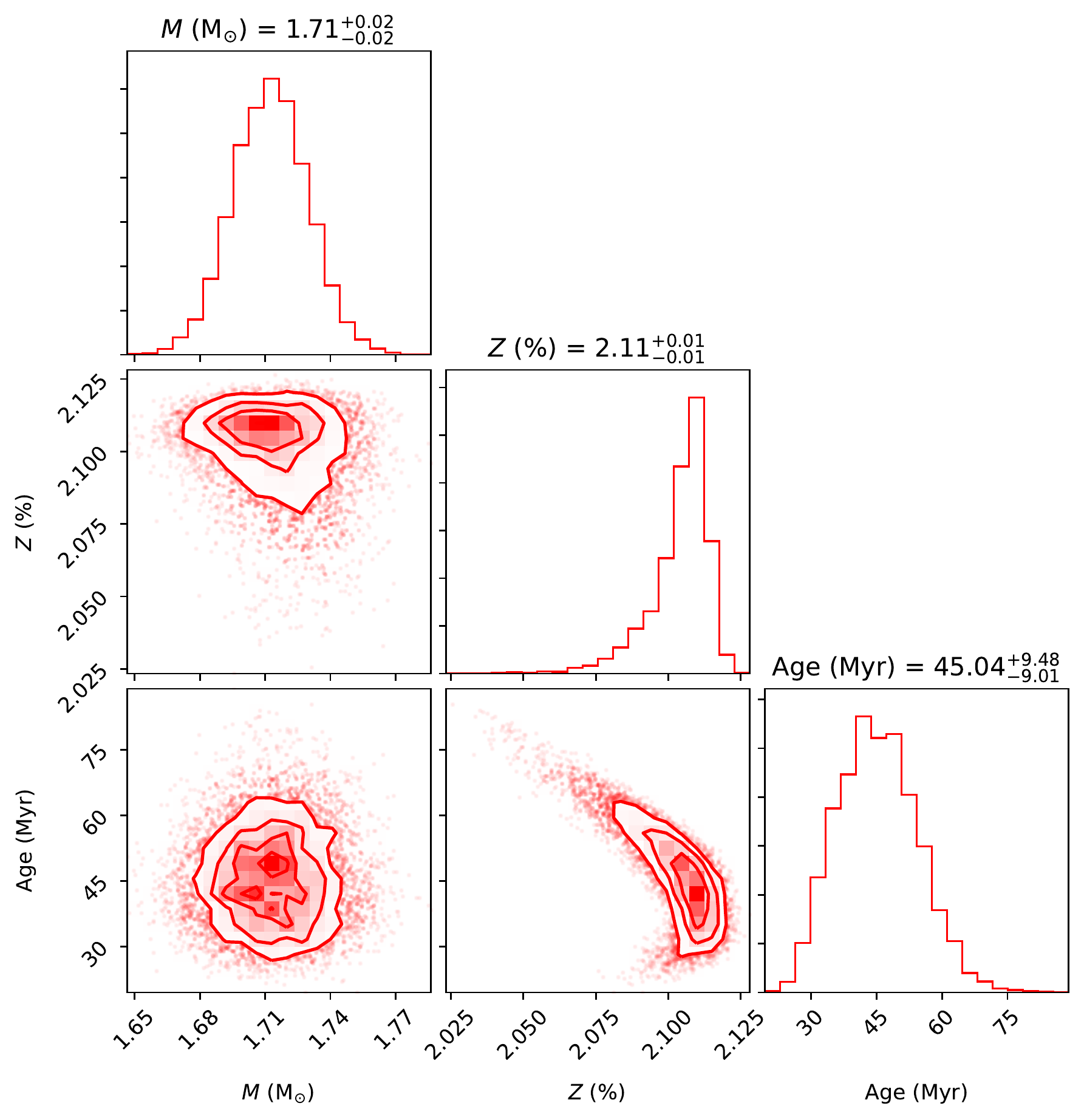}
\put(45,74){\large HD\,31901}
\end{overpic}
\begin{overpic}[abs,unit=1mm,scale=0.48]{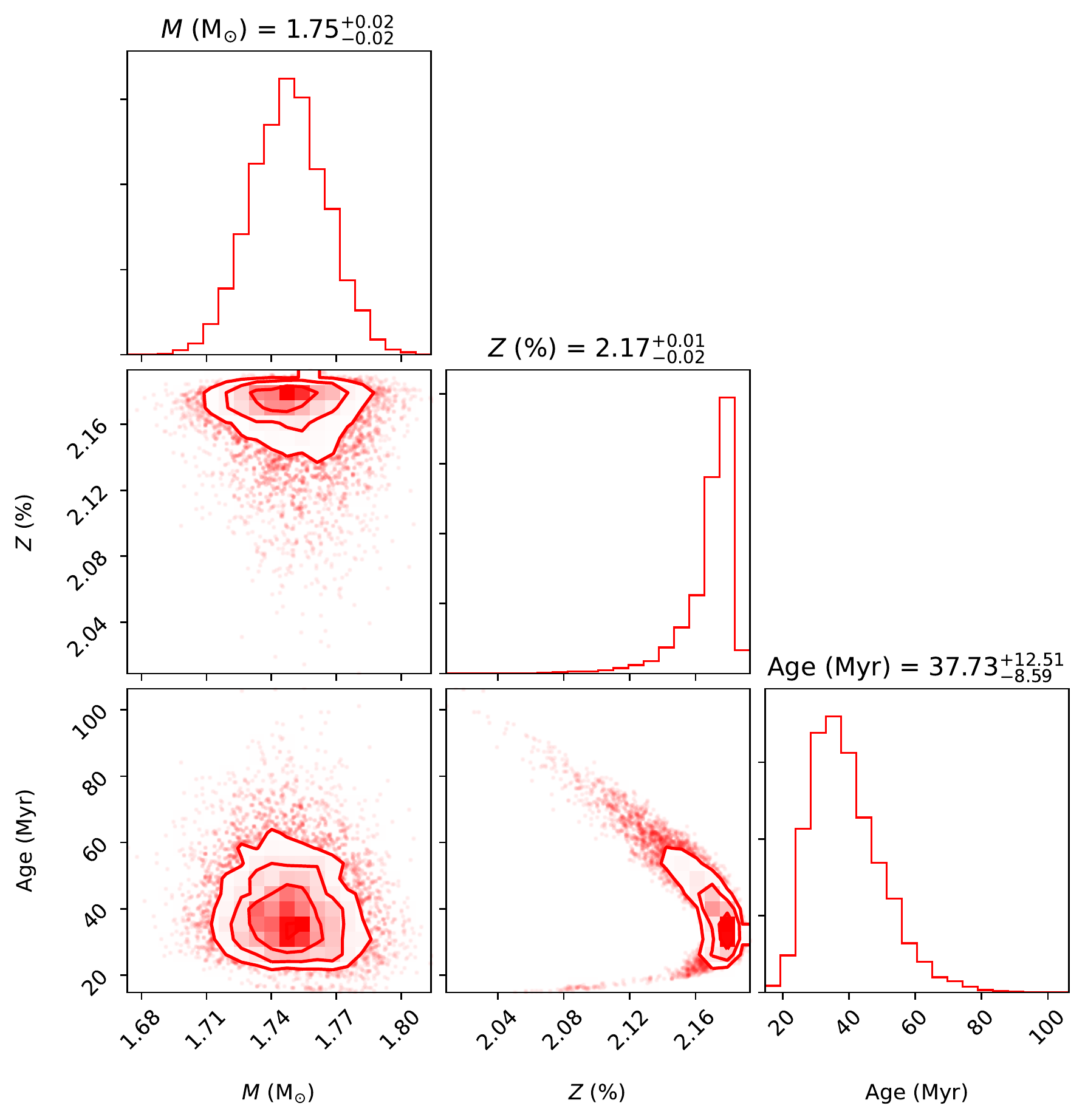}
\put(45,74){\large HD\,55863}
\end{overpic}
\caption{Corner plots for the analysis of two of the three remodelled stars from \citet{beddingetal2020}: HD\,31901 ($33\pm4$\,km\,s$^{-1}$; top), HD\,55863 ($v\sin i = 99\pm5$\,km\,s$^{-1}$; bottom).}
\label{fig:corners1}
\end{center}
\end{figure}

\begin{figure}
\begin{center}
\begin{overpic}[abs,unit=1mm,scale=0.48]{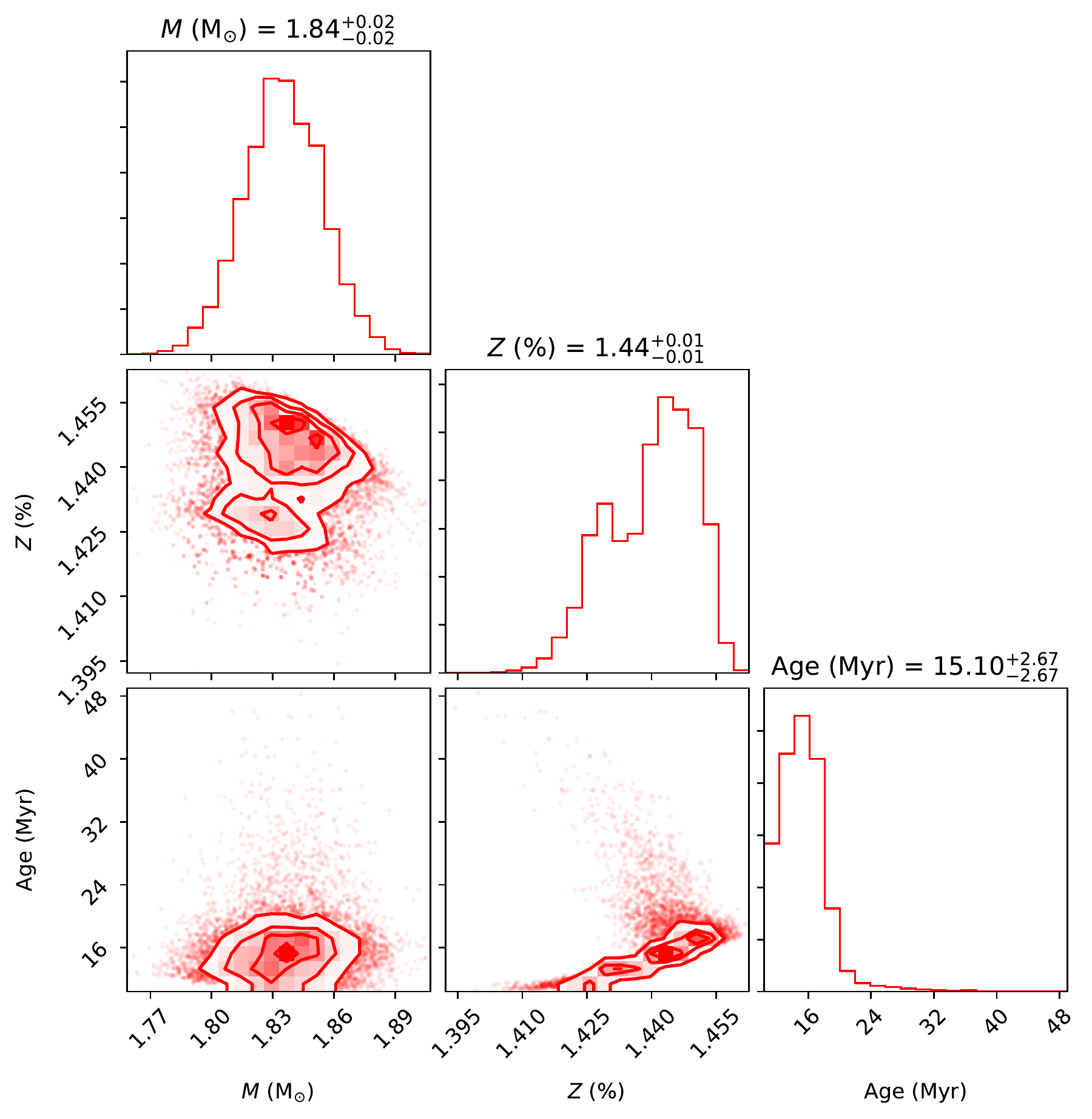}
\put(45,74){\large HD\,28548}
\end{overpic}
\begin{overpic}[abs,unit=1mm,scale=0.48]{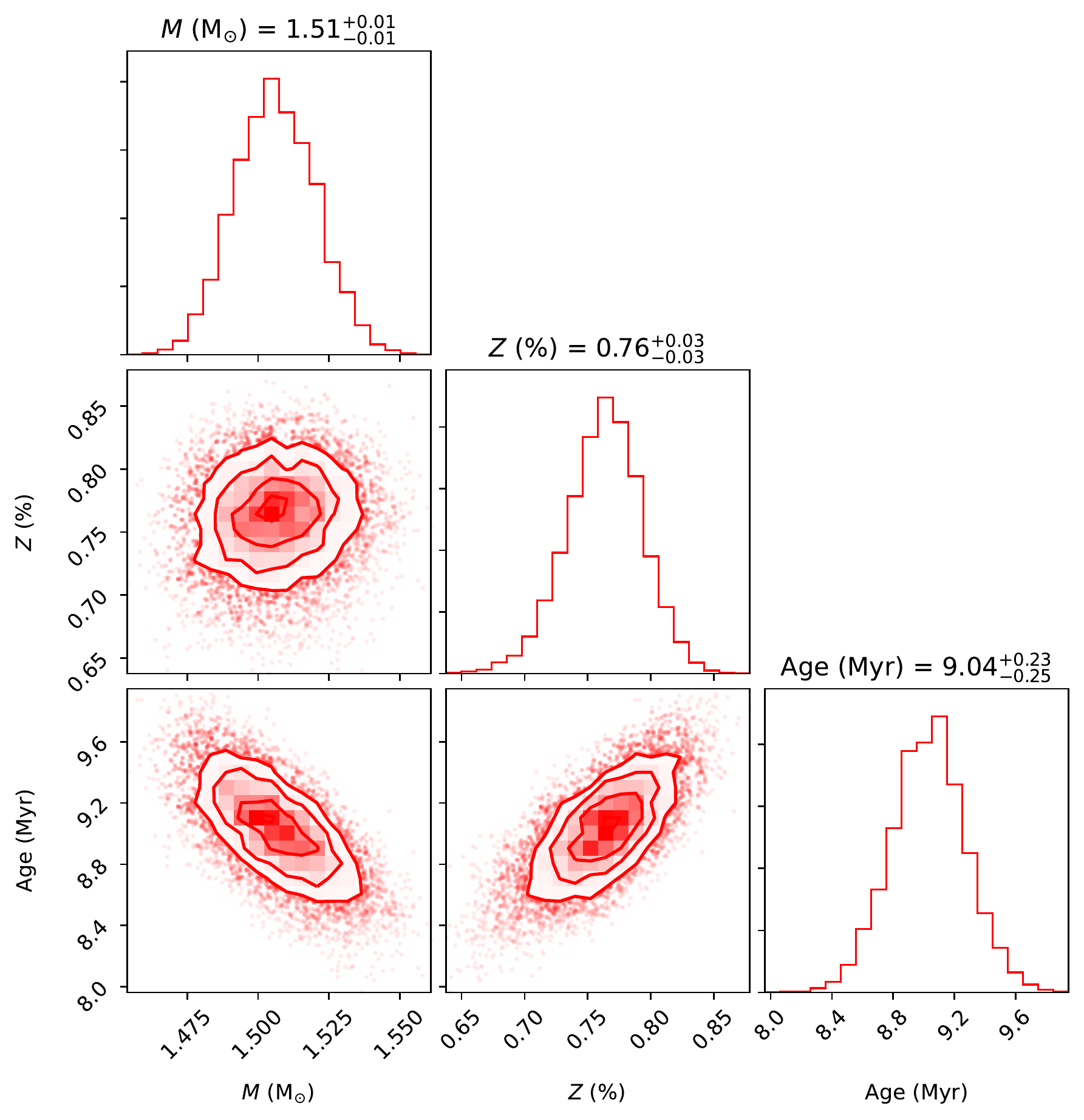}
\put(45,74){\large HD\,46722}
\end{overpic}
\caption{Corner plot for HD\,28548 ($v\sin i = 200\pm50$\,km\,s$^{-1}$; top), and for HD\,46722 (bottom).}
\label{fig:corners2}
\end{center}
\end{figure}

\begin{figure}
\begin{center}

\begin{overpic}[abs,unit=1mm,scale=0.48]{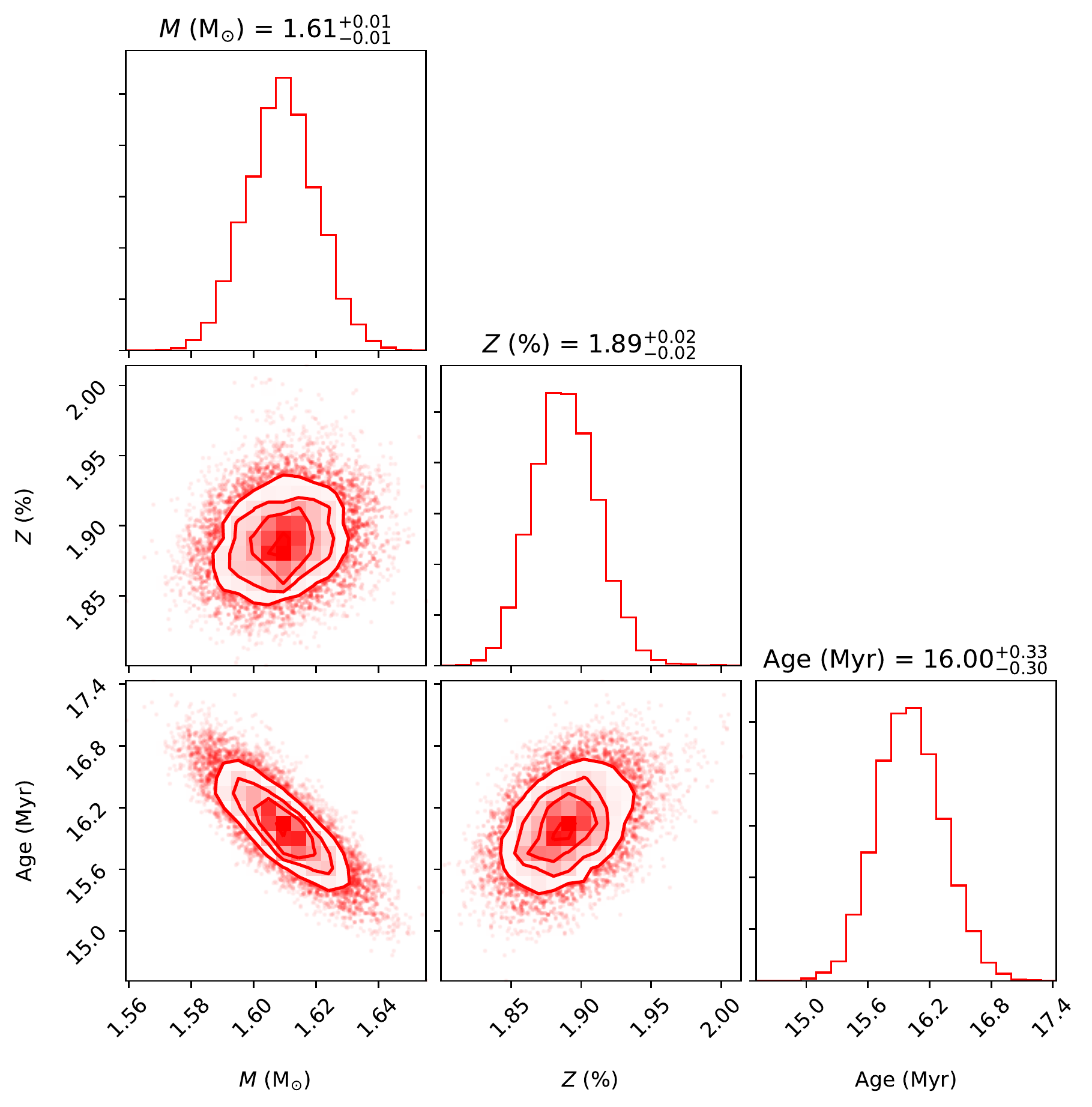}
\put(43,74){\large HD\,112063}
\end{overpic}
\caption{Corner plot for HD\,112063.}
\label{fig:corners3}
\end{center}
\end{figure}

\begin{figure*}
\begin{center}
\includegraphics[width=0.33\textwidth]{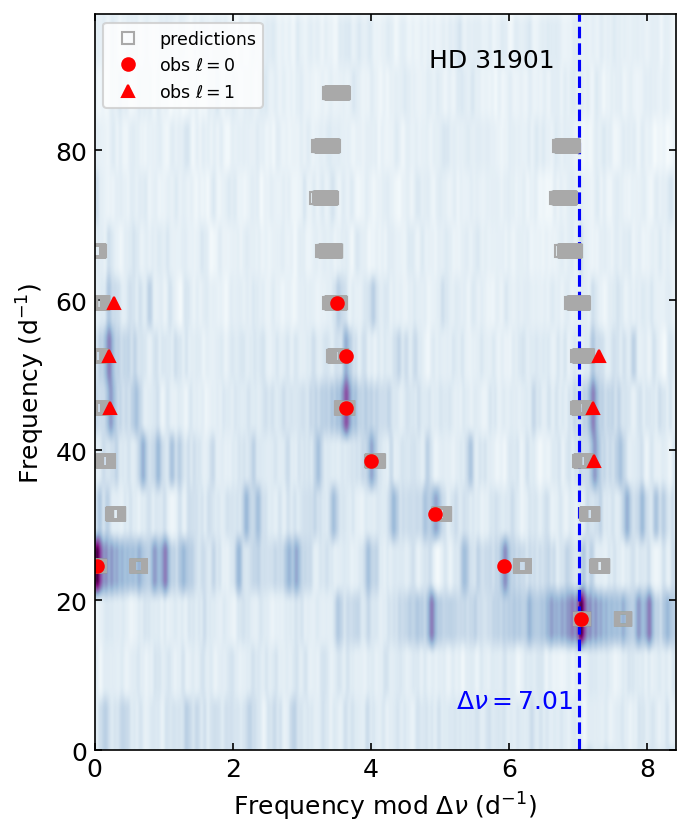}
\includegraphics[width=0.33\textwidth]{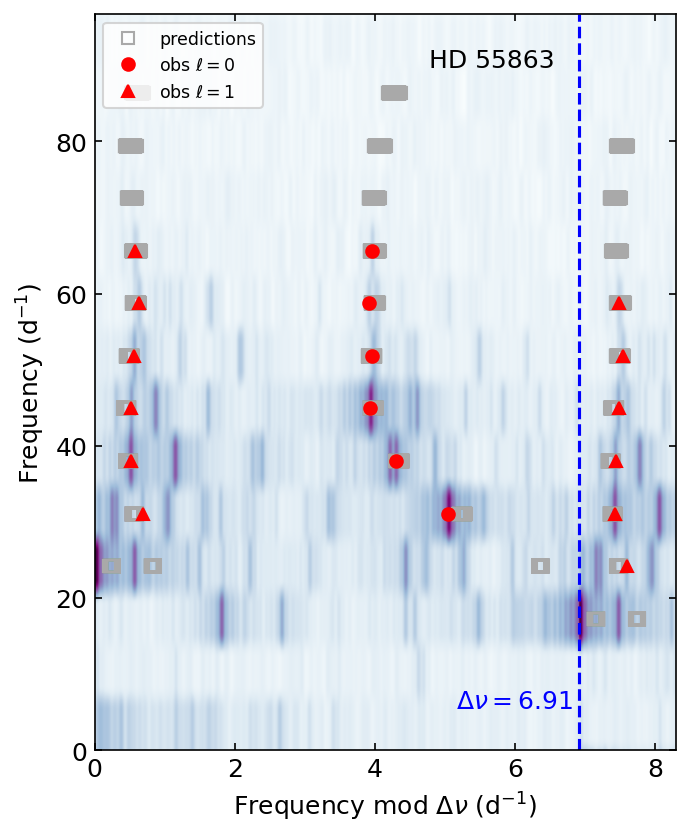}
\includegraphics[width=0.33\textwidth]{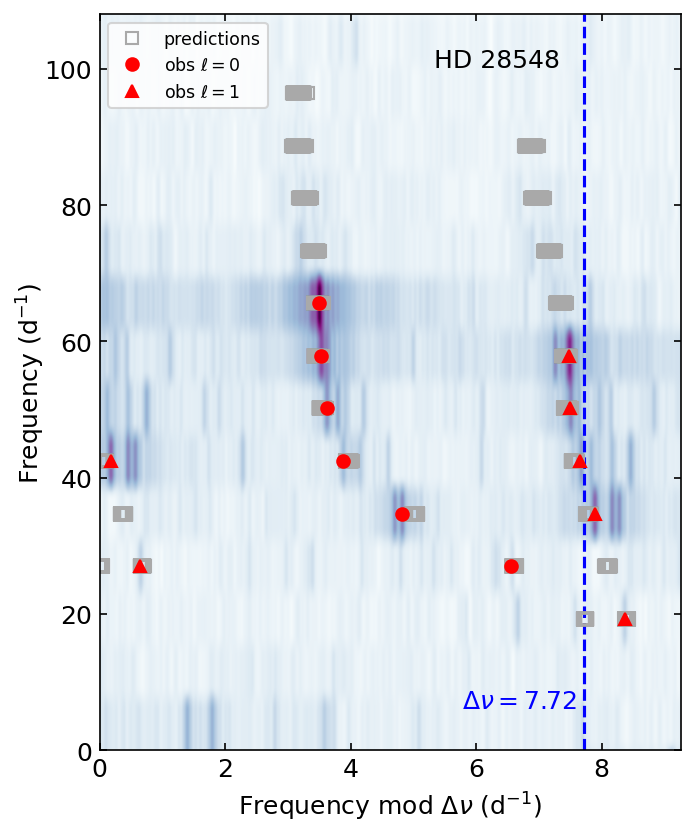}\\
\includegraphics[width=0.33\textwidth]{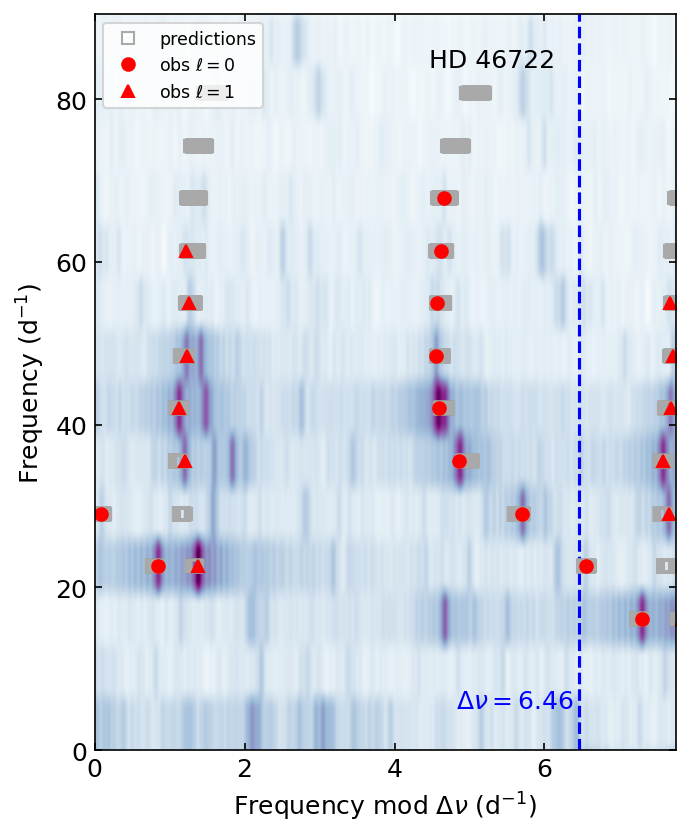}\includegraphics[width=0.33\textwidth]{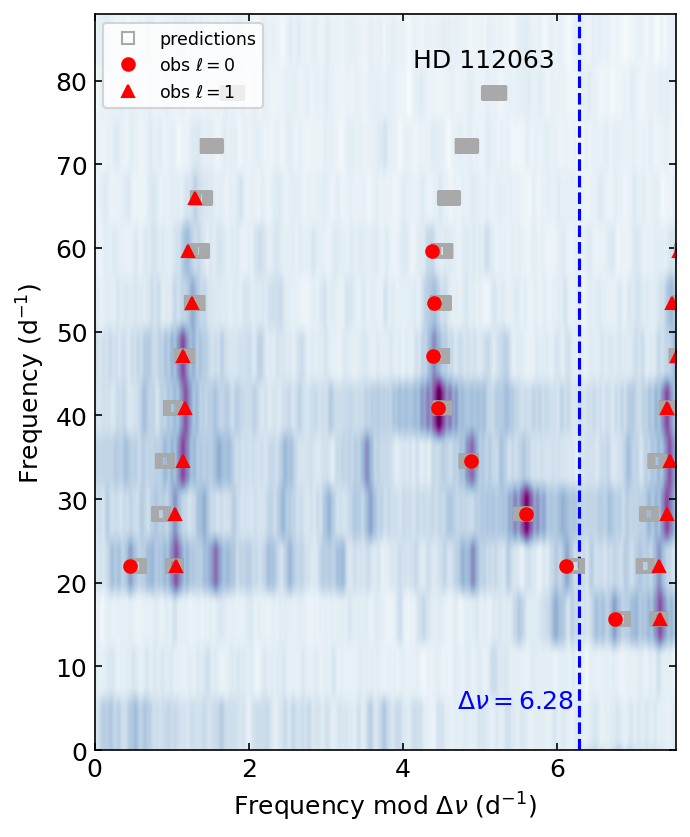}\\
\caption{\'Echelle diagrams of five $\delta$\,Sct stars. The grey-scale shows the observed amplitude spectra; the red points mark the identified modes, and the overlapping grey squares show posterior-predicted mode frequencies from the neural network.}
\label{fig:pred_ech1}
\end{center}
\end{figure*}

\bsp	
\label{lastpage}
\end{document}